\documentclass{article}

\usepackage{arxiv}

\usepackage[utf8]{inputenc} % allow utf-8 input
\usepackage[T1]{fontenc}    % use 8-bit T1 fonts
\usepackage{hyperref}       % hyperlinks
\usepackage{url}            % simple URL typesetting
\usepackage{booktabs}       % professional-quality tables
\usepackage{amsfonts}       % blackboard math symbols
\usepackage{nicefrac}       % compact symbols for 1/2, etc.
\usepackage{microtype}      % microtypography
\usepackage{lipsum}
\usepackage{graphicx}
\graphicspath{ {./images/} }
\usepackage{makecell}
\usepackage{tabularx}
\usepackage[table]{xcolor}
\usepackage{circledsteps}
\usepackage{enumitem}
\usepackage{soul}
\usepackage[labelfont=bf,textfont=it]{caption}
\usepackage[labelfont=bf,textfont=it]{subcaption}
\setul{0.35ex}{0.2ex}
\usepackage{calc}
\usepackage{array}
\usepackage{placeins}
\usepackage{amsmath}

\title{Investigating Performance and Practices with Univariate Distribution Charts}

\author{
 Laura Lotteraner \\
  University of Vienna\\
  Vienna, Austria \\
  \texttt{laura.lotteraner@univie.ac.at} \\
   \And
 Anna Kurtenkova \\
  University of Vienna\\
  Vienna, Austria \\
  \And
 Torsten Möller \\
  University of Vienna\\
  Vienna, Austria \\
 \And
 Daniel Pahr \\
  University of Vienna\\
  Vienna, Austria \\
}

\definecolor{boxplotblue}{RGB}{41, 125, 115}
\definecolor{violinplotgreen}{RGB}{119, 137, 41}
\definecolor{stripplotyellow}{RGB}{234, 163, 22}
\definecolor{histogrampink}{RGB}{142, 39, 96}

\definecolor{sig}{RGB}{34, 131, 187}
\definecolor{nonsig}{RGB}{185, 217, 235}

\newcommand{\Boxplot}{\Circled[inner color=boxplotblue, outer color = boxplotblue]{\sffamily \textbf{\textcolor{boxplotblue}{B}}}}
\newcommand{\Violinplot}{\Circled[inner color=violinplotgreen, outer color = violinplotgreen]{\sffamily \textbf{\textcolor{violinplotgreen}{V}}}}
\newcommand{\Stripplot}{\Circled[inner color=stripplotyellow, outer color = stripplotyellow]{\sffamily \textbf{\textcolor{stripplotyellow}{S}}}}
\newcommand{\Histogram}{\Circled[inner color=histogrampink, outer color = histogrampink]{\sffamily \textbf{\textcolor{histogrampink}{H}}}}
 
\newcommand{\Mean}{\setulcolor{black}{\sffamily\textsc{Mean}}}
\newcommand{\Median}{\setulcolor{black}{\sffamily\textsc{Median}}}
\newcommand{\Range}{\setulcolor{black}{\sffamily\textsc{Range}}}
\newcommand{\Clusters}{\setulcolor{black}{\sffamily\textsc{Clusters}}}
\newcommand{\Describe}{\setulcolor{black}{\sffamily\textsc{Describe}}}

\newcolumntype{Y}{>{\centering\arraybackslash}p{0.7cm}}

\newcommand{\good}{\textcolor{blue}{\pmb{\Large $+$}}}
\newcommand{\neutral}{\textcolor{gray!20!black}{\pmb{\Large $\sim$}}}
\newcommand{\bad}{\textcolor{red!90!black}{\pmb{\Large $-$}}}

\begin{document}
\maketitle

\begin{figure}[!h]
    \centering
    \includegraphics[width=\linewidth]{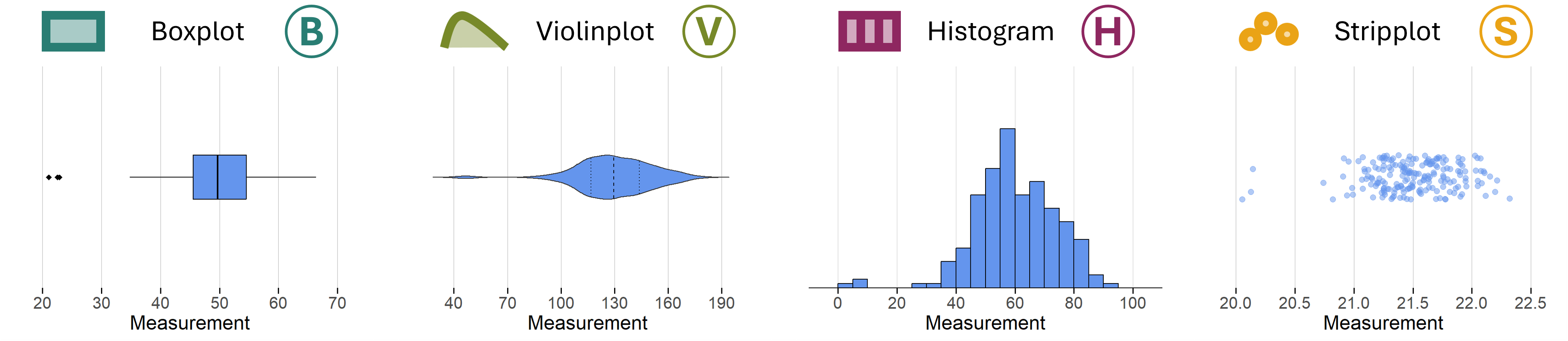}
    \caption{The four charts used in this study (dataset 1):  \emph{\Boxplot~\textbf{\textcolor{boxplotblue}{Boxplot}}, \Violinplot~\textbf{\textcolor{violinplotgreen}{Violinplot}}, \Histogram~\textbf{\textcolor{histogrampink}{Histogram}}}, and  \emph{\Stripplot~\textbf{\textcolor{stripplotyellow}{Stripplot}}}, each representing one of the four classes  of charts: \includegraphics[height=1em]{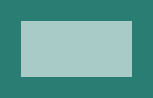} Summary Statistics, \includegraphics[height=1em]{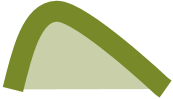} Smooth Densities, \includegraphics[height=1em]{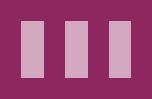} Binned Densities, \includegraphics[height=1em]{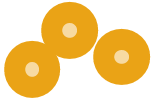} Individual Data Points. }
    \label{fig:charts_overview}
\end{figure}

\begin{abstract}
%why 1d charts important
A range of charts with different strengths and weaknesses exists to support the visual analysis of univariate distributions, with a limited understanding of which charts best support which tasks and users, and how practitioners use charts.
%what we do
We categorize the available charts for univariate distributions into four groups and present the results of a mixed-methods comparison (n=215) of participants' perception and preferences across boxplots, violinplots, jittered stripplots, and histograms as representatives of their respective categories. The click-to-select approach in our study, combined with data on participants' subjective experiences and preferences, allows to both measure accuracy on benchmark tasks and discuss participants' choices qualitatively.

%Outcome
Our analysis reveals differences between charts in task accuracy, common misunderstandings, and preferences across various low-level tasks, and indicates that chart preference and familiarity do not necessarily align with participants’ task performance.
Interviews with five visualization practitioners further reveal that charts widely preferred by general audiences (such as histograms) or commonly used in scientific domains (such as boxplots) are not inherently the most effective for all tasks.
\end{abstract}

\section{Introduction}

% What is the problem?
Numerous charts have been developed to display various characteristics of univariate data distributions, including boxplots~\cite{tukey_exploratory_1977} and histograms. Objectively, these charts have clear advantages and disadvantages. Boxplots, for example, do not reveal any information about the distribution's shape~\cite{matejka_same_2017}, but are ideal for a quick comparison of summary statistics. Histograms and smooth density curves, on the other hand, are ideal for assessing the distribution's shape but lack an explicit encoding of summary statistics. 
Charts displaying the individual data points, by contrast, get cluttered and fail to convey information when the dataset is too large.

Even a chart appropriate for a task, however, is of little use if the intended audience struggles to interpret it or is unfamiliar with it. This underscores the importance of studying how people interpret various charts and identifying which visualization approaches domain experts adopt in practice.
% Why is it interesting and important?
Investigating both questions together is interesting, as a gap might exist between which charts are easy to interpret and which are actually used. 
Both the use of suboptimal charts out of convenience and the misinterpretation of objectively well-suited charts can conceal important information in the data, which is relevant for both data analysis and the communication of results.

% Why is it hard? (E.g., why do naive approaches fail?)
Assessing how people interpret different charts presents challenges. Even a simple chart, such as a boxplot, has many variations~\cite{mcgill_variations_1978} that can affect understanding, so choosing appropriate variants is a crucial part of any study. Furthermore, the actual design of a chart, including color, orientation, scale, and aspect ratio, gridlines, annotations, and other embellishments, may also have an influence~\cite{skau_evaluation_2015}. Similarly, results are valid only for the tasks they were tested on, and many possible tasks exist~\cite{blumenschein_v-plots_2020}.
Various studies have assessed~\cite{lem_misinterpretation_2013, lem_experts_2014} and compared~\cite{rodrigues_comparing_2019, blumenschein_v-plots_2020} univariate distribution charts  with respect to both preference and accuracy in various tasks. Others~\cite{sahann_histogram_2021,correll_looks_2019} took a closer look at the impact of the data and chart features.
All of these previous studies offer valuable insights by comparing examples from the design space. However, the influence of the type of chart on both performance and user experience has not been systematically investigated using both representative charts and a diverse participant sample.

% What are the key components of my approach and results? 
In this study, we contribute to closing this gap by conducting a mixed-methods study with both lay and expert audiences.
Our research questions are \textbf{(RQ-1)}: \textit{How do different types of univariate distribution charts influence participants' performance and experience?}
and \textbf{(RQ-2)}: \textit{What are common practices and preferences in using these charts for data analysis?}
To answer these questions, we present the results of a survey with a diverse audience to assess participant performance, experience, and preferences. Our charts (see Figure~\ref{fig:charts_overview}) were selected to cover different aspects of the wide range of univariate distribution charts, hereafter referred to as "distribution charts". To address RQ-2, we further contrast our study results with interviews with five researchers from various domains who use charts in their daily work.
Our contributions are
\begin{itemize}
    \item[-] a selection of charts from four distinct groups of distribution charts,
    \item[-] a mixed methods study (n = 215) on participant performance, experience, and preferences in benchmark tasks with these charts,
    \item[-] semi-structured interviews with five domain experts on the application and experience with visualizing univariate distributions.
\end{itemize}

By triangulating our findings, we show that chart preference and familiarity do not necessarily align with participants' chart performance and discuss how this gap can be addressed in future work.

The code used to create our stimuli, as well as the collected survey responses, and an appendix with additional figures, are available at \href{https://osf.io/7r6cx/overview?view_only=ddde1c1417634736957075912a79ce3a}{OSF}.
%-------------------------------------------------------------------------

\section{Background and Related Work}
We present both charts and tasks relevant to our study, along with findings from other studies on the performance of these charts. 

\subsection{Charts for Univariate Data Distributions}\label{subsec:related_charts}

We group charts for univariate data distributions into five categories based on the representations presented by Kay~\cite{kay_ggdist_2024}, who maps functions of distributions, such as densities or quantiles, onto aesthetics to visualize uncertainty. Of the four types of representations Kay distinguishes, we consider those applicable to our data: intervals, slabs, and dotplots. Importantly, we discuss the display of observed data, not theoretical distributions.

\paragraph*{\includegraphics[height=1em]{B.png} Summary Statistics.} One concrete example of Kay's~\cite{kay_ggdist_2024} proposed \textit{interval representations} are classic ``error bars''~\cite{correll_error_2014, hullman_hypothetical_2015}. These representations display selected descriptive statistics of a distribution, and are thus closely related to the boxplot~\cite{tukey_exploratory_1977}, its predecessors  
\textit{range-bar charts}~\cite{haemer_range-bar_1948}, and \textit{range plots}~\cite{spear_charting_1952}, and other related charts. 
Boxplot variants include variable-width boxplots~\cite{mcgill_variations_1978} for different sample sizes, notched boxplots~\cite{mcgill_variations_1978}, letter-value plots~\cite{hofmann_letter-value_2017} for large data, adjusted boxplots for skewed distributions~\cite{hubert_adjusted_2008}, the box-less plot~\cite{tufte_envisioning_1990}, a colored variant~\cite{carr_colorful_1994}, and both varying definitions of quantiles and values of $k$ for the length of the whiskers ($k \cdot \textnormal{IQR}$)~\cite{frigge_implementations_1989}. 

\paragraph*{\includegraphics[height=1em]{V.png} Smooth Densities.} Smooth density curves are one instance of Kay's~\cite{kay_ggdist_2024} proposed \textit{slab representations}, which show the distribution density. Typically, smooth density curves display values on the horizontal axis~\cite{fygenson_impact_2025} and differ primarily in the kernel density estimators used to calculate them. They can be combined into a ridgeline plot~\cite{liu_ridgebuilder_2025} to compare several data distributions. Violinplots~\cite{hintze_violin_1998}, on the other hand, are defined by a mirrored (i.e., symmetric) density curve, with values usually on the vertical axis. While violinplots often come with the median and the quartiles explicitly marked, many design variants exist~\cite{molina_how_2022}, and their only common characteristic is the symmetric density curve. Instead of height (density curve) or width (violinplot), gradient plots use opacity to encode the probability density function~\cite{kay_when_2016}.

\paragraph*{\includegraphics[height=1em]{H.png} Binned Densities.} Distribution densities can not only be displayed as smooth curves, but also in aggregation by predefined value bins, most commonly in the form of histograms. No consensus exists on an optimal bin count~\cite{sahann_histogram_2021}, and while variable bin widths may better reflect the distribution~\cite{heim_accustripes_2024}, the data range is often divided into uniform bins.  Other binned charts include accustripes~\cite{heim_accustripes_2024}, stem-leaf diagrams~\cite{wilkinson_grammar_2005}, histodot plots~\cite{wilkinson_dot_1999}, and stacked histograms for two-dimensional data~\cite{knittel_visual_2021}. 

\paragraph*{\includegraphics[height=1em]{S.png}  Data Points.} The classic dotplot~\cite{wilkinson_dot_1999} stacks individual data points, with more recent adaptations accounting for large frequency differences~\cite{rodrigues_nonlinear_2018}.  Alternatives to stacking are random jittering, structured layout algorithms~\cite{sidiropoulos_sinaplot_2018, eklund_beeswarm_2010, rodrigues_relaxed_2023}, and the display along a single line, often as stripes instead of dots~\cite{leisch_neighborhood_2010}. Displaying individual points is most effective for small sample sizes~\cite{correll_looks_2019}.

\paragraph*{\includegraphics[height=1em]{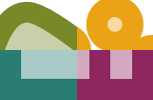} Hybrid Charts.} Finally, we encountered several charts that are hybrids of two or more charts from the previously discussed categories. Hybrid charts are often considered ``defensive'' in the sense that the individual components ``defend'' against situations where any single one of them might lead to data misinterpretation~\cite{correll_teru_2023}. Such hybrid charts include the  vaseplot~\cite{benjamini_opening_1988}, the histplot~\cite{benjamini_opening_1988}, the sea stack plot~\cite{stuart_sea_2024}, the beanplot~\cite{kampstra_beanplot_2008}, the rugplot~\cite{Yau2012_VisualizeCompareDistributions}, the v-plot~\cite{blumenschein_v-plots_2020}, the summary plot~\cite{ potter_visualizing_2010} and the raincloud plot~\cite{allen_raincloud_2021, correll_teru_2023}.

\subsection{Assessment and Comparison of Distribution Charts}
Many studies have evaluated the effectiveness of individual charts for specific audiences and tasks. Studies report difficulties in interpreting boxplots and histograms among first-year university students and experts alike~\cite{lem_misinterpretation_2013, lem_experts_2014}, and show that design parameters such as the bin size in a histogram can affect distribution understanding~\cite{sahann_histogram_2021}. More broadly, different representations may conceal certain data features, indicating that no single visualization design ``will make all potential data quality issues equally visible''~\cite{correll_looks_2019}.

Controlled comparisons further demonstrate task-dependent differences between chart types. For example, boxplots have been reported to underperform histograms for distribution characterization and anomaly detection~\cite{rodrigues_comparing_2019}, yet to perform well for mean identification in normal distributions~\cite{newburger_fitting_2023}. Other work shows that distributional encodings influence effect size and confidence judgment~\cite{kale_visual_2021, helske_can_2021}, the interpretation of uncertainty (e.g., reliability judgments~\cite{fernandes_uncertainty_2018, gschwandtner_visual_2016, correll_error_2014} and hypothetical outcomes~\cite{hullman_hypothetical_2015, kale_hypothetical_2019}), and the comparison of two distributions~\cite{newburger_comparing_2023}. Complementing performance-based evaluations, preference studies suggest that no single visualization is considered universally useful across analysis tasks~\cite{blumenschein_v-plots_2020, helske_can_2021}.

Domain-specific investigations in transportation and environmental decision-making likewise demonstrate that effectiveness depends on context and comparison setting~\cite{kay_when_2016, kozlova_visual_2020, fernandes_uncertainty_2018, riedel_replacing_2022}.

Most studies on participants' task performance with visualizations of univariate distributions are focusing on specific charts~\cite{sahann_histogram_2021, lem_experts_2014}, audiences~\cite{rodrigues_comparing_2019, lem_misinterpretation_2013}, and tasks~\cite{newburger_comparing_2023, kozlova_visual_2020}, and comparison between studies is difficult due to variations in study and chart design. Building on previous work, we compare one representative from each of the basic chart categories introduced above with a general audience (n=215), aiming to examine central trade-offs within the design space of visualizations of univariate distributions. Our study investigates participants’ perception of these charts through free exploration~\cite{north_toward_2006}, benchmark tasks~\cite{quadri_survey_2022}, and click-based behavioral analysis~\cite{kim_bubbleview_2017}.

\subsection{Tasks for Univariate Data Distributions}
Insights about data visualizations depend on the tasks used to evaluate them. Numerous tasks, spanning different levels of Bloom’s taxonomy~\cite{bloom_taxonomy_1956}, have been proposed, and extensive theoretical work has classified and structured these tasks.

Shneiderman~\cite{shneiderman_eyes_1996} proposes a data-type-by-task taxonomy of visualizations, comprising seven high-level tasks. Building on this, Amar et al.~\cite{amar_low-level_2005} present a set of ten low-level analysis tasks that capture most activities involved in studying data through data visualizations. North~\cite{north_toward_2006} makes the distinction between benchmark tasks, which identify specific low-level effects, and open-ended questions to identify insights participants gain from a visualization. He emphasizes the importance of asking open-ended questions before any benchmark tasks. Brehmer and Munzner~\cite{brehmer_multi-level_2013} provide an abstraction of domain-specific into concrete visual analysis tasks.  

In previous studies comparing charts for univariate data distributions, tasks at different abstraction levels and levels of Bloom's taxonomy~\cite{bloom_taxonomy_1956} have been used. Lem et al.~\cite{lem_misinterpretation_2013, lem_experts_2014} embedded their low-level tasks (\textit{compare sample sizes}, \textit{compare sample means}, \textit{identify minimum value}, \textit{find specific value}, \textit{interpret the median}) in short scenarios that were relatable to their audience. All questions were phrased as either binary (true/false) or single-choice items with predefined response options. Similarly, Rodrigues et al.~\cite{rodrigues_comparing_2019}, who worked with a real-world dataset, used single-choice questions sourced from a pilot study for their low-level tasks. Only two of their tasks, \textit{characterize distribution} and \textit{find anomalies}, were tested on charts relevant to our study. Blumenschein et al.~\cite{blumenschein_v-plots_2020} distinguish between local (including \textit{identify the frequency of a value}), aggregation (including \textit{identify sample mean}), and global tasks (including \textit{describe shape of distribution}) and categorize tasks by type (identification vs. comparison) and complexity (on a scale from 1 to 4). They did not test performance on these tasks; however, they asked for the perceived usefulness of different charts. 

While ``characterizing the distribution'' is often phrased as an open question~\cite{north_toward_2006}, in some studies it is presented as a single-choice question~\cite{sahann_histogram_2021} or a lineup protocol. In \textit{lineup protocols}~\cite{correll_looks_2019, vanderplas_spatial_2016}, participants are asked to detect the one visualization among a ``lineup'' of \textit{n} visualizations that looks ``different'' or has a specific feature. 

\section{Survey Design}
To answer research question \textbf{RQ-1}, we investigated the performance of the four chart types for distribution characterization, summary statistics, and comparison tasks in a crowdsourced survey. We piloted the survey with 145 students to test the chart representatives and the overall survey procedure.

\subsection{Charts}

From the chart categories for univariate data distributions outlined in Section~\ref{subsec:related_charts}, we selected one commonly used representative per category (see Figure~\ref{fig:charts_overview}).
All charts were created using R~\cite{R_Core_2021} and the \texttt{ggplot} package~\cite{ggplot}, a widely used tool among (expert and non-expert) chart creators. Because design choices within a chart can influence performance and preferences~\cite{skau_evaluation_2015}, we relied on default settings wherever possible to reflect common practice. Below, we explain the rationale for each of the four charts.

\begin{figure*}[ht!]
    \centering
    \includegraphics[width=\linewidth]{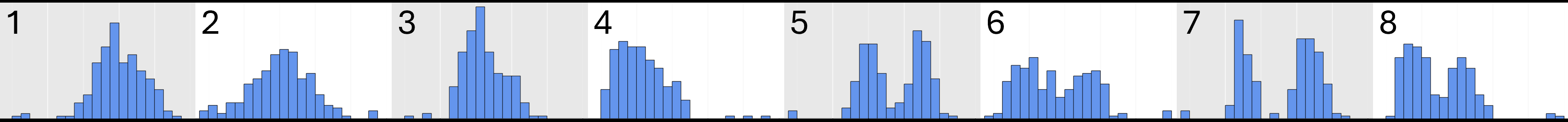}
    \caption{The datasets used in the study vary in skew (symmetric: 1, 2, 5, 6; left-skewed: 3, 4, 7, 8), outlier values (left: 1, 3, 5, 7; right: 2, 4, 6, 8), and modality (unimodal: 1–4; bimodal: 5–8).}
    \label{fig:datasets}
\end{figure*}

\paragraph*{\includegraphics[height=1em]{B.png}} The
\Boxplot~\textbf{boxplot} (\texttt{geom\_boxplot}) in our survey is based on the original publication by Tukey~\cite{tukey_exploratory_1977}, with the whiskers representing the quartiles $\pm 1.5 \cdot \textnormal{IQR}$, and any values beyond that range individually plotted as outliers. By using this standard boxplot, we aim to gain insight into how people understand the fundamental features of a boxplot.

\paragraph*{\includegraphics[height=1em]{V.png}}  While horizontal density curves are the simplest representative of smooth density charts, we chose the
\Violinplot~\textbf{violinplot} (\texttt{geom\_violin}) for several reasons. Fewer of the previous studies compared the violinplot~\cite{hullman_hypothetical_2015, gschwandtner_visual_2016, correll_error_2014, helske_can_2021, riedel_replacing_2022} to other charts than horizontal density curves~\cite{gschwandtner_visual_2016, correll_looks_2019, fernandes_uncertainty_2018, kale_visual_2021, kay_when_2016, newburger_comparing_2023}, so less is known about how people perform on it. Furthermore, it has been shown~\cite{correll_error_2014} that in certain situations, symmetric charts are less confusing than asymmetric ones. By using the symmetric violinplot, we can link the errors made to the smooth curve rather than the asymmetry. We also aimed to balance the explicit display of summary statistics in the charts, and thus chose a violinplot with explicitly drawn median and quartiles, which is highly unusual for horizontal density curves.

\paragraph*{\includegraphics[height=1em]{H.png}} For binned data, we selected the \Histogram~\textbf{histogram} (\texttt{geom\_hist}) as the canonical and most widely used representation of aggregated univariate data distributions. By removing the vertical axis, we make the histogram comparable to the other three chart types.

\paragraph*{\includegraphics[height=1em]{S.png}} For the display of individual data points, we used a simple jittered 
\Stripplot~\textbf{stripplot} (\texttt{geom\_jitter} with \texttt{alpha = 0.5}). This approach provides a minimal representation of individual observations while mitigating overplotting through jitter and partial transparency, without adding visual encodings such as size or shape variation. Despite using dots rather than stripes, we call this plot ``stripplot'', to distinguish it from stacked dotplots~\cite{wilkinson_dot_1999} and related variants~\cite{rodrigues_nonlinear_2018, rodrigues_relaxed_2023}, which introduce such additional encodings.

Initially, we used the charts in their most common orientation, which means that values are on the horizontal axis for histograms and on the vertical axis for all other charts. Due to confusion during our pilot survey, however, we opted to rotate the boxplot, violinplot, and jittered stripplot to align with the histogram's orientation. We rotated these plots rather than the histogram to preserve the histogram’s familiar appearance. The pilot survey showed that participants were generally familiar with histograms but not with charts such as violinplots, making orientation changes less problematic for the latter. To minimize the influence of design choices within a chart, colors, labels, grids, and ticks were held constant across charts.

\subsection{Data}

Following Correll et al.~\cite{correll_looks_2019}, we generated synthetic data. To ensure that our results were unrelated to the specific dataset used in the survey, we created a total of eight datasets, %differing in \textbf{skew}, \textbf{outlier values}, and \textbf{modality}, 
see Figure~\ref{fig:datasets}. We distinguish between symmetric data sampled from a normal distribution and left-skewed data sampled from a beta distribution. To reduce the number of datasets in the study and because we assumed the direction of skew would not influence the results, we excluded right-skewed datasets. Three outlier values, that is, values outside the whiskers of a boxplot, were included either on the left or the right side. The full code for dataset generation is available in our supplemental material.

Each dataset was then scaled differently for each of the four chart types. Similar to Correll et al.~\cite{correll_looks_2019}, we selected a fixed dataset size of 200 data points. Based on the 50 data points they used to generate unimodal distributions, we increased the size to allow for sufficient display of bimodal distributions and of groups of outlier values that are still clearly distinguishable from data clusters. Overlap effects in the jittered stripplot remain similar to those observed by Correll et al.\cite{correll_looks_2019} due to the vertical jitter we introduce. For each task and chart type, participants saw a random dataset to further avoid learning effects from one chart to the next. 

\subsection{Tasks}

We based our tasks both on the \textit{low-level components of analytic activity in information visualization} by Amar et al.~\cite{amar_low-level_2005} and the comprehensive list of tasks compiled by Blumenschein et al.~\cite{blumenschein_v-plots_2020}. From their lists of tasks, we selected those relevant to both univariate data distributions and our intended general audience (see Table~\ref {tab:Tasks}). We focus on a combination of free exploration, followed by low-level retrieval and binary comparison tasks without a specific domain context to narrow the focus of our study on participants' understanding of the visualizations.

\paragraph*{Free Exploration.} 
Following the recommendation of North~\cite{north_toward_2006}, each participant was asked to freely \Describe~one of the eight datasets for each chart prior to the benchmark tasks.
Without a specific goal or time limit, participants could mark up to ten interesting points in the chart by clicking on their corresponding locations and provided a written description of points of interest in a free-text field.

\begin{table*}[!h]
\captionsetup{justification=raggedright,singlelinecheck=false}
\caption{Tasks used in this study and how they relate to Blumenschein et al.~\cite{blumenschein_v-plots_2020} and Amar et al.~\cite{amar_low-level_2005}. }
\label{tab:Tasks}
\centering
\begin{tabularx}{\textwidth}{lXp{0.48\textwidth}p{0.18\textwidth}}
\toprule
\textbf{Task Code} & \textbf{Task Name} & \textbf{Blumenschein et al.}~\cite{blumenschein_v-plots_2020} & \textbf{Amar et al.}~\cite{amar_low-level_2005} \\
\midrule
\Mean~& Identify Mean & identify the mean of one distribution  & retrieve value \\
 & &   & compute derived value   \\
  & Compare Mean & compare the means of multiple distributions  & compare two datasets \\
\rowcolor{gray!20} \Median~& Identify Median & identify the median of one distribution & retrieve value  \\
\rowcolor{gray!20} & & & compute derived value   \\
\rowcolor{gray!20}  & Compare Median & compare the medians of multiple distributions & compare two datasets \\
\Range~& Identify Range &   & find extremum\\
&  &   & determine range \\
\rowcolor{gray!20} \Clusters~& Identify Clusters & compare frequencies within one distribution &   \\
 \Describe~& Describe Dataset & describe and identify the shape and type of one distribution  & characterize distribution  \\
& & describe and identify the skewness and kurtosis of one distribution  &    \\
\bottomrule
\end{tabularx}
\end{table*}

\paragraph*{Benchmark Tasks.}
We chose the identification or estimation of the \Median~to evaluate the efficiency of the explicit encodings used in boxplots and violinplots, and the
\Mean~as a value more sensitive to extreme outliers and not typically explicitly encoded in summary statistics charts.
To test perception of data clusters, we asked participants to identify \Clusters~in unimodal and bimodal datasets. Because the boxplot does not show clusters, we provided participants with a way to specify that the task cannot be solved.
Finally, we had participants identify the \Range~of values in a distribution, by explicitly selecting the minimum and maximum values in the chart, which implicitly requires participants to reason about the existence of outliers.
To limit overall study time, we implemented the \Median~and \Mean~tasks as comparisons, requiring participants to select the relevant value \textit{in} the distribution where it is higher.

\subsection{Procedure}
We employed a within-subjects design, with all participants completing all five tasks for each of the four charts. The study was implemented in QuestionPro~\cite{questionpro} and consisted of several parts.

\paragraph*{Demographics. } Due to a known influence of domain background on performance at visualization tasks~\cite{hall_professional_2022}, we collected participant demographics, including age, education level, and field of expertise, to account for these effects in our analysis.

\paragraph*{Pre Survey.} Before the start of the actual study, participants received onboarding~\cite{stoiber_visualization_2019} to the purpose of the study and to the four chart types, and three practice tasks to familiarize themselves with the click-to-select approach. Due to the elementary nature of the survey tasks and our interest specifically in \textit{how} participants (mis)interpret the charts, we did not provide the actual survey tasks for training. The participants then rated their knowledge of and experience with \textit{Statistics}, \textit{Data Visualization}, and how often they \textit{saw}, \textit{interpreted} and \textit{created} each of the four charts. 

\paragraph*{Main Survey.} 

For each task, participants first received an explanation of the relevant statistical concept. They were then shown a random dataset for each of the four charts in random order and, for each chart, completed the task, rated their confidence, and assessed perceived difficulty. After evaluating all four charts, they ranked them by perceived usefulness and explained their ranking. All participants saw \Describe~ first to avoid priming; the remaining tasks (\Median, \Range, \Mean, \Clusters) were presented in counter-balanced order. To filter out randomly clicking participants, each task block included one attention check requiring the selection of a specified single-choice item. Due to our interest in misinterpretations, we deliberately did not check task understanding. 

\paragraph*{Post Survey.} At the end of the study, participants rated both their level of comfort using each chart during the study and their likelihood of using the charts in the future.

\subsection{Metrics}
We collected both qualitative and quantitative results. 

\paragraph*{Qualitative Data.}

We coded each free-text chart description by a participant in the \textbf{free exploration task} by the distribution attributes mentioned, namely summary statistics, distribution shape or symmetry, as well as points and areas of interest such as clusters, modes and outliers. Obvious misunderstandings (e.g., interpreting histogram bins as time points) were marked as misreads.
Two co-authors independently coded each description. After the initial coding round, we created a unified codebook and performed a second coding round for validation, achieving an inter-coder agreement of $92\%$ for $1,650$ assignments. We resolved the remaining conflicts in a final discussion.
We split the statements justifying \textbf{participants' chart rankings} into $1,471$ separate utterances, each pertaining to exactly one chart and task. 
We then collectively developed a codebook to categorize the utterances into one of 10 types, grouped into three categories: appeal, visual encoding, and usability.
Afterwards, a single co-author classified the utterances according to the codebook, additionally assigning positive or negative valence.

\paragraph*{Quantitative Data.} Benchmark tasks were evaluated as follows:
\begin{enumerate}
    \item For solving a task, participants were asked to click on the chart where they believed the correct value to be, or into a designated area below the chart if they did not believe the task could be solved using the chart(s). Only clicks on the chart are counted as \textit{valid}. Because identifying \Clusters~in \Boxplot~boxplots is not possible, the definition is reversed in this case.
    \item For \Mean~and \Median, two charts were presented above each other, with the ``invalid'' area below the lower chart. The \textit{comparison} task is counted as correct if the click is on the correct chart, and the click is assigned to the chart it was placed on.
    \item For \Range~and \Clusters, in which more than one click was possible, each click is then assigned to the correct value closest to it, and the \textit{accuracy} is calculated as the absolute error normalized by the range of the respective dataset to make values comparable across the differently scaled charts. 
    \item For each correct value, the click with the lowest relative error is assigned the \textit{type} ``closest'', all others are assigned the type ``additional''.
    \item If no click is assigned to a correct value, it is marked as \textit{missing}.
\end{enumerate}

\begin{figure*}[t!]
    \centering
    \begin{subfigure}[t]{0.24\textwidth}
        \centering
        \includegraphics[width=.99\textwidth]{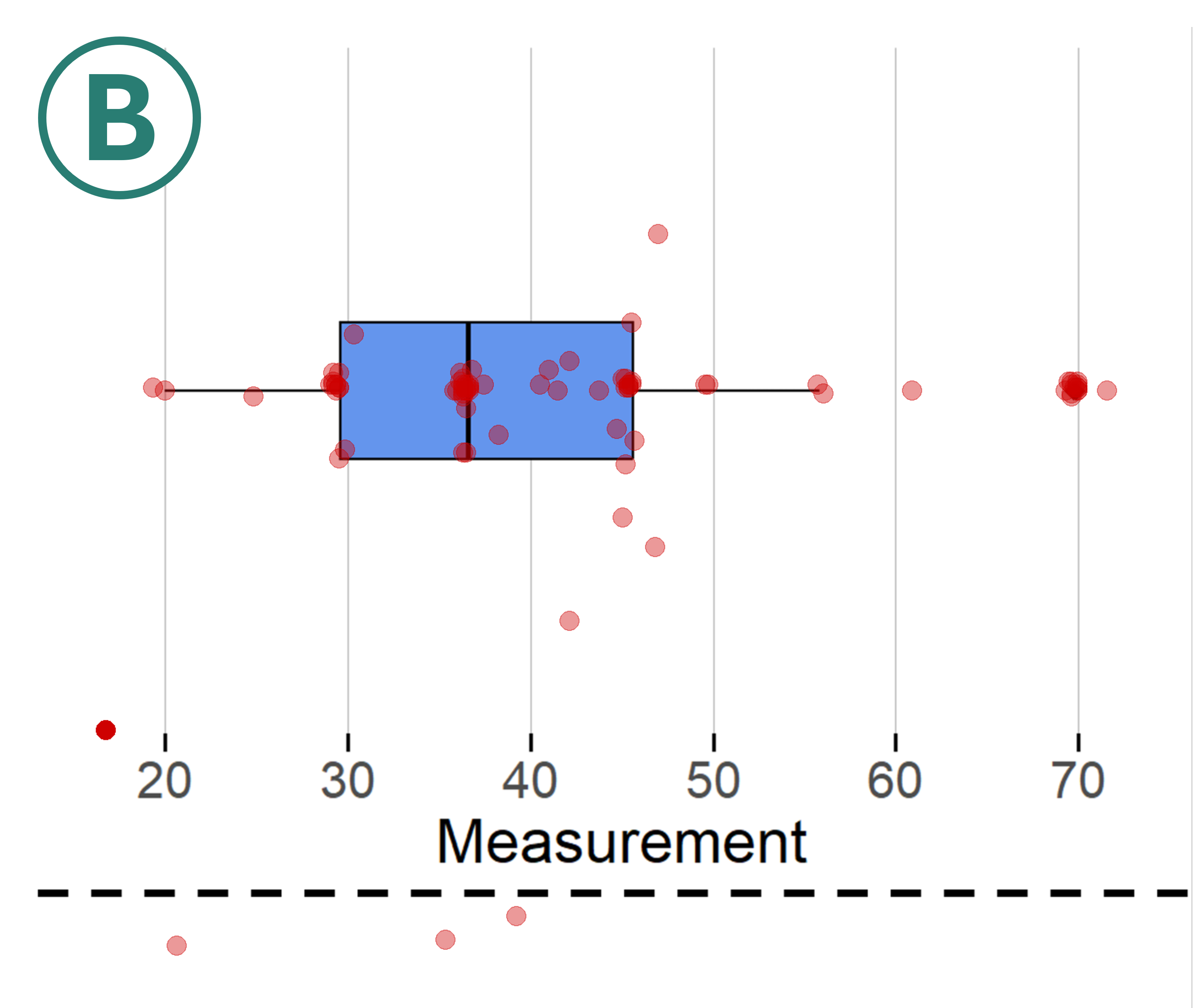}
    \end{subfigure}%
    ~
    \begin{subfigure}[t]{0.24\textwidth}
        \centering
        \includegraphics[width=.99\textwidth]{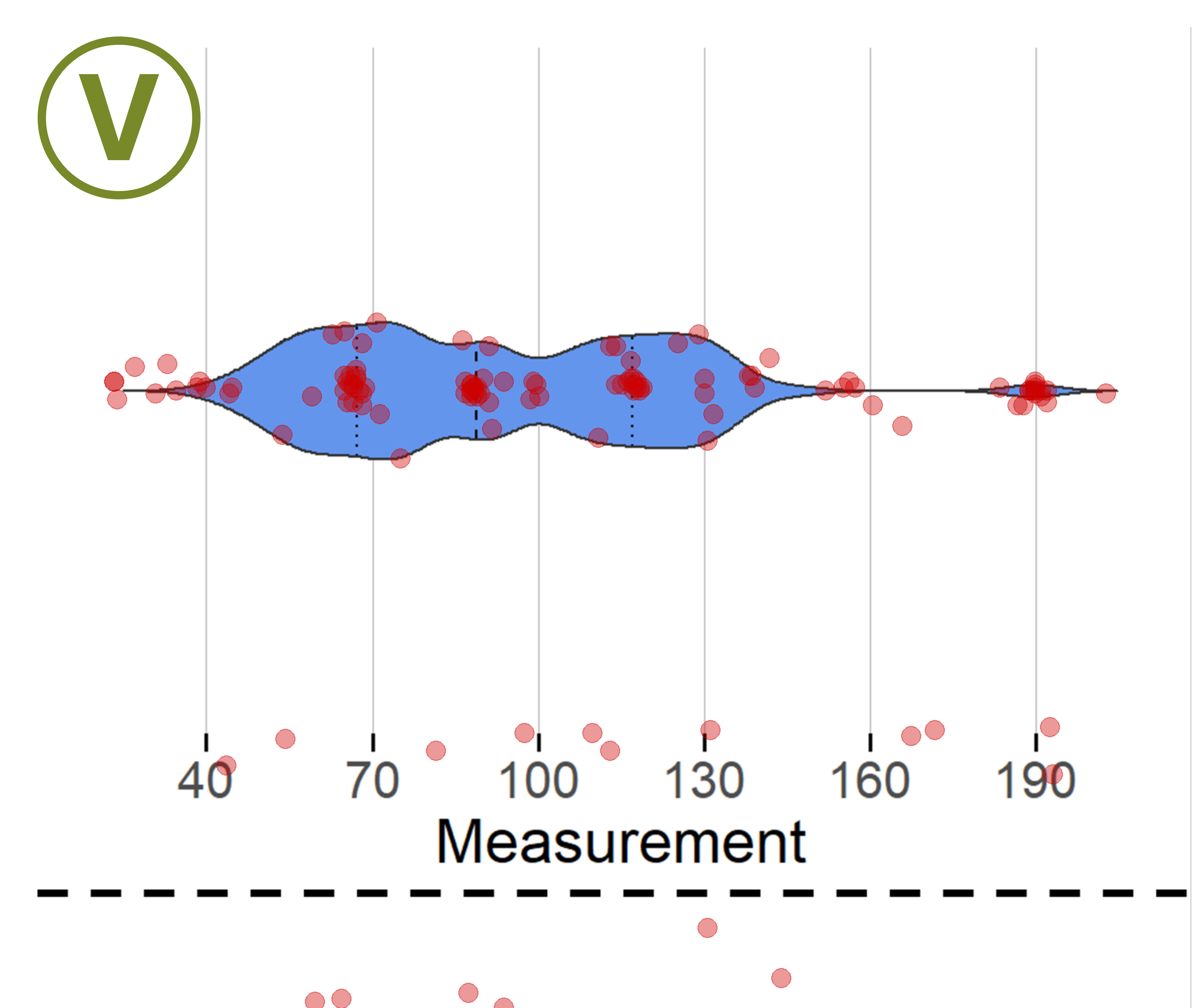}
    \end{subfigure}
    ~
    \begin{subfigure}[t]{0.24\textwidth}
        \centering
        \includegraphics[width=.99\textwidth]{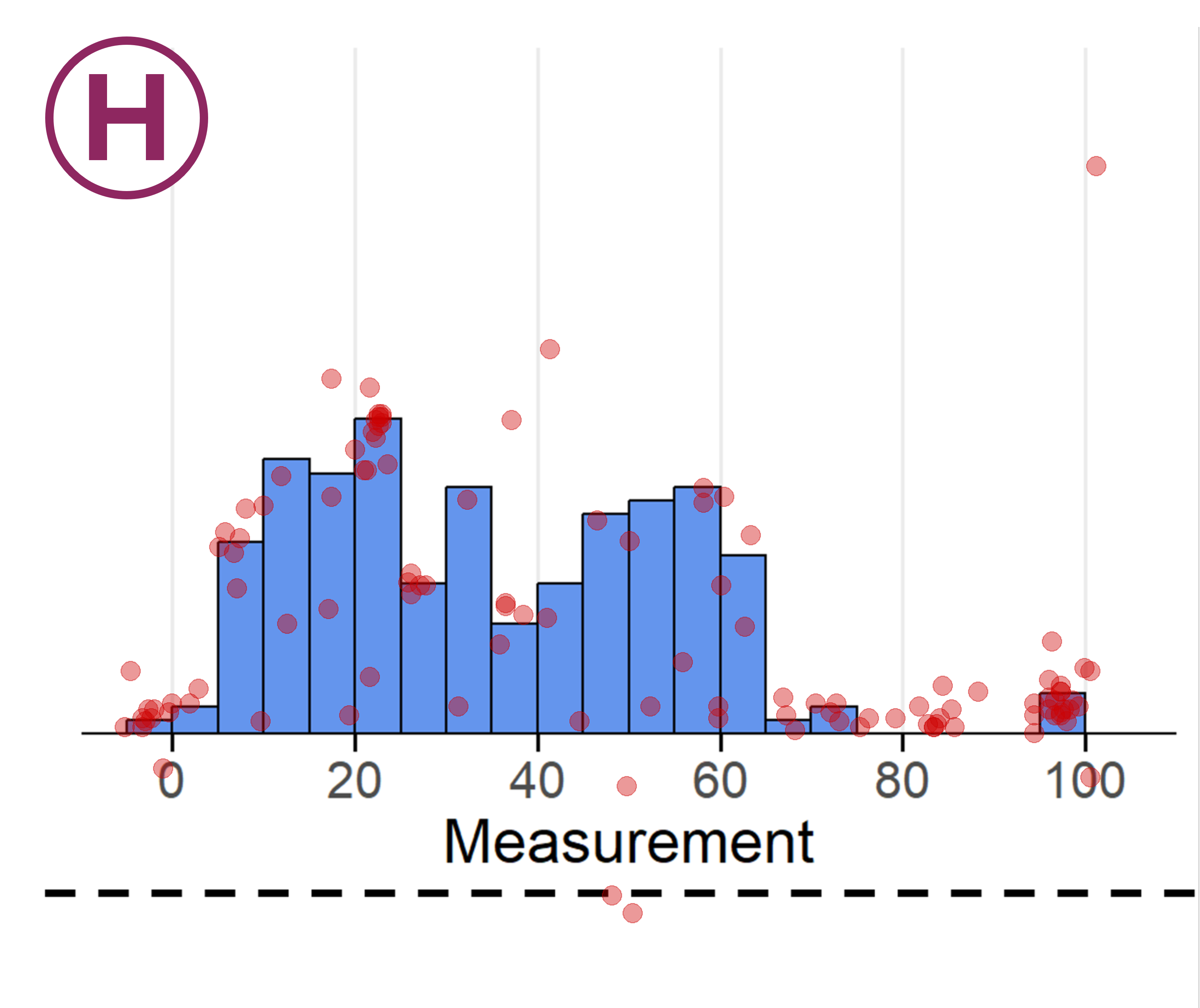}
    \end{subfigure}
    ~
    \begin{subfigure}[t]{0.24\textwidth}
        \centering
        \includegraphics[width=.99\textwidth]{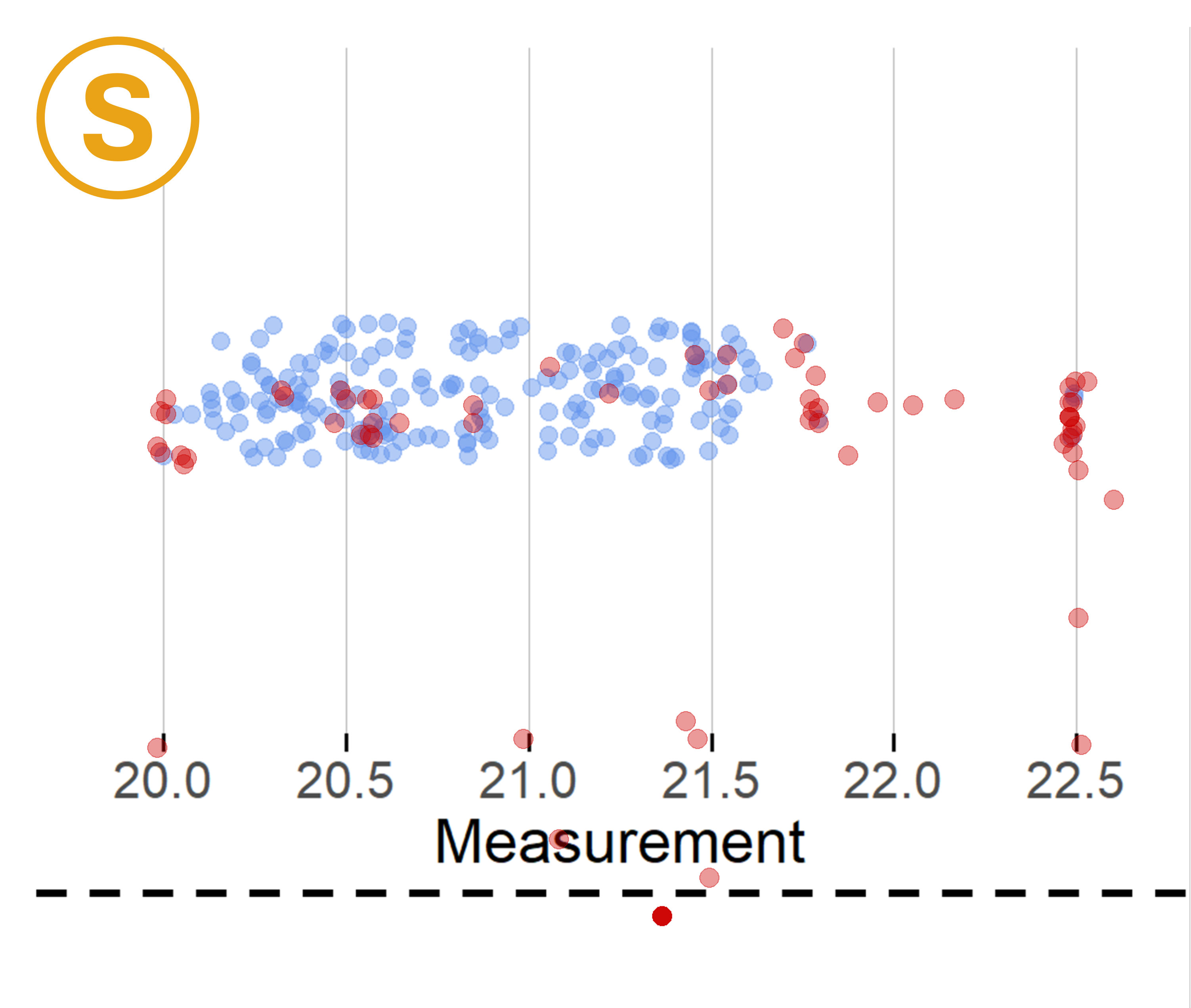}
    \end{subfigure}
    \caption{Examples of participants' selections on dataset 6 in the \Describe~task, for which participants could leave up to 10 marks on a chart to denote points of interest. A red dot represents a mark left by a participant. \emph{\Boxplot}~Boxplots exhibit a high concentration on the median line and around outliers. \emph{\Violinplot}~Violinplots show similar behavior also on quartile markings, as well as peaks and valleys. \emph{\Histogram}~Histograms are marked predominantly at bar height. Jittered \emph{\Stripplot}~stripplots tend to be marked at high point concentrations. 
    %. 
    }\label{fig:charts_clickmap}
\end{figure*}

\textbf{Confidence} and perceived \textbf{difficulty} were measured on a 5-point scale, \textbf{preferences} after each task by ranking the charts from most (rank = 1) to least (rank = 4) preferred, and overall preferences for both during the survey and future use on a 5-point scale per chart.

\subsection{Recruitment}
%A power analysis based on the primary effect size from our pilot survey indicated a minimum sample size of 200. 
Following a power analysis based on the primary effect size from our pilot survey, and to ensure a participant pool including both a general audience and experts, we recruited 200 participants via 
Prolific~\cite{douglas_data_2023} (over 18 and fluent in English), and experts from various domains through network-based and snowball sampling. The participants on Prolific were paid an average of \pounds6.85 per hour, and exceptionally fast submissions were excluded automatically.

\section{Survey Results}
In this section, we discuss the results of the online survey.  

\subsection{Participants}
Excluding participants who failed more than one of the four attention checks resulted in 215 survey participants whose responses were used in the analysis, aged between 18 and 75 years. 106 participants identified as female, 105 as male, 3 as non-binary, and one did not specify their gender. Participants' highest level of education ranged from no higher degree (89) to a graduate degree (52). Major fields of work included computer science and engineering (86), humanities or social sciences (36), healthcare (34), natural and life sciences (31), business and finance (25), and education (23). Full participant demographics are provided in the appendix.

For each of the four charts, the three familiarity items (\textit{See}, \textit{Interpret}, \textit{Create}), reported on a 5-point scale, showed high internal consistency (Cronbach’s $\alpha > .70$) and were thus averaged to one \textit{Familiarity} score per chart with mean values 1.6 (violinplot), 2.0 (boxplot), 2.3 (jittered stripplot) and 4.0 (histogram). For the domains statistics and data visualization, both \textit{Knowledge} and \textit{Experience}, reported on a 5-point scale, showed high internal consistency (Cronbach's $\alpha = 0.87$) and were thus averaged to one \textit{Knowledge} score per domain, with mean values 2.8 (Statistics) and 3.0 (Data Visualization). Full details can be found in the appendix. 
 
\subsection{Qualitative Results}

First, we discuss participant behavior in selecting points of interest in the four different plots at hand of the data collected in the \Describe~task.
We summarize the relative occurrences of insights from the free exploration task per attribute, the misreads per chart, and the relative occurrences of codes per chart, as reported in the participants' written feedback. We provide absolute numbers (N) and relative occurrences (in \%) of codes in brackets, as well as example quotes in cursive. We excluded the statements from a single participant due to their obvious use of AI tools to generate their answers. 
For the remaining open feedback, we present a brief summary of our exploratory analysis.
Common mistakes in the benchmark tasks are discussed through a visual representation of the participant selection behavior compared to the ground truth.
For a detailed breakdown of the qualitative results, we refer to the appendix.

\begin{figure*}[t!]
    \centering
    \begin{subfigure}[t]{0.235\textwidth}
        \centering
        \includegraphics[width=.99\textwidth]{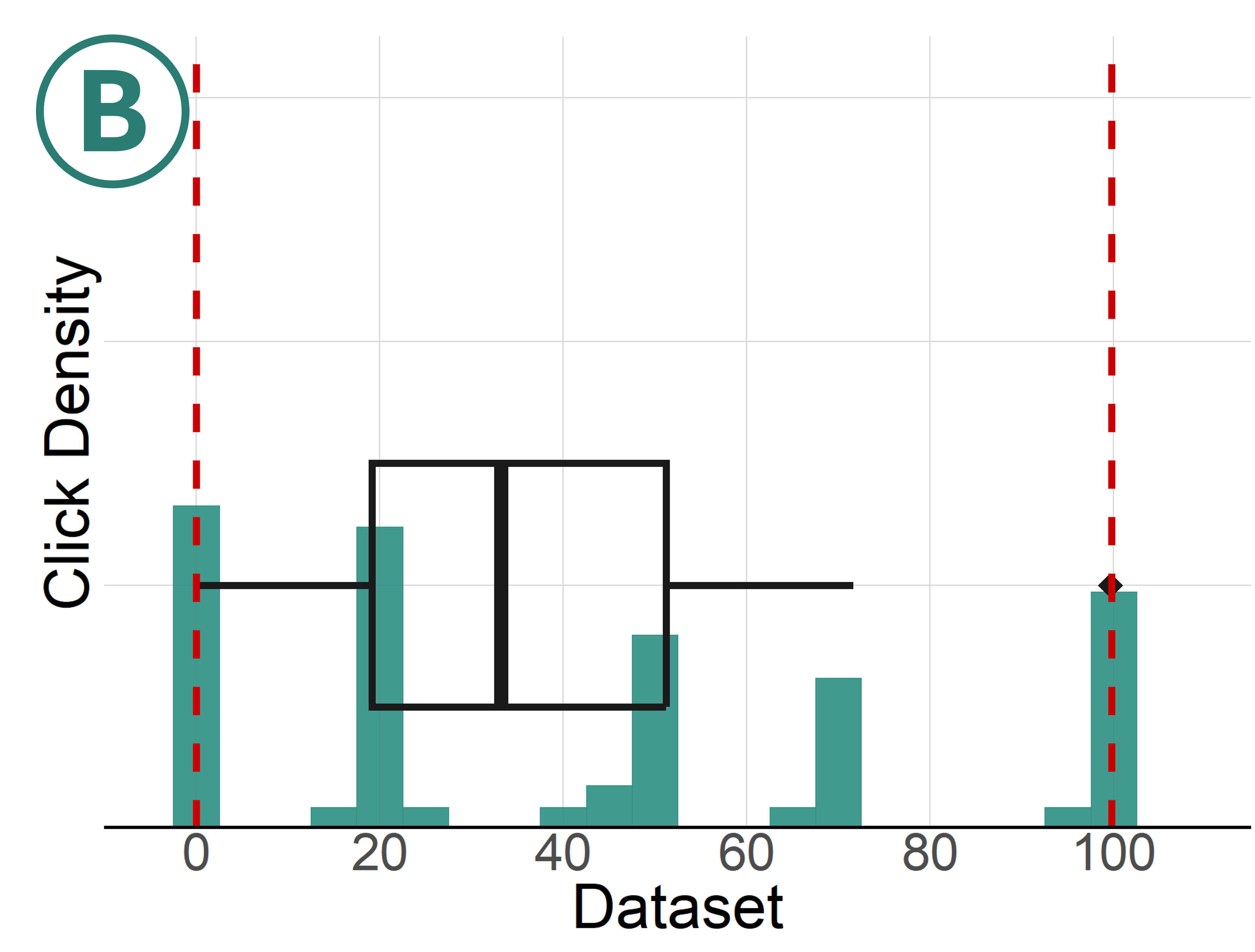}
    \end{subfigure}
        ~ 
    \begin{subfigure}[t]{0.235\textwidth}
        \centering
        \includegraphics[width=.99\textwidth]{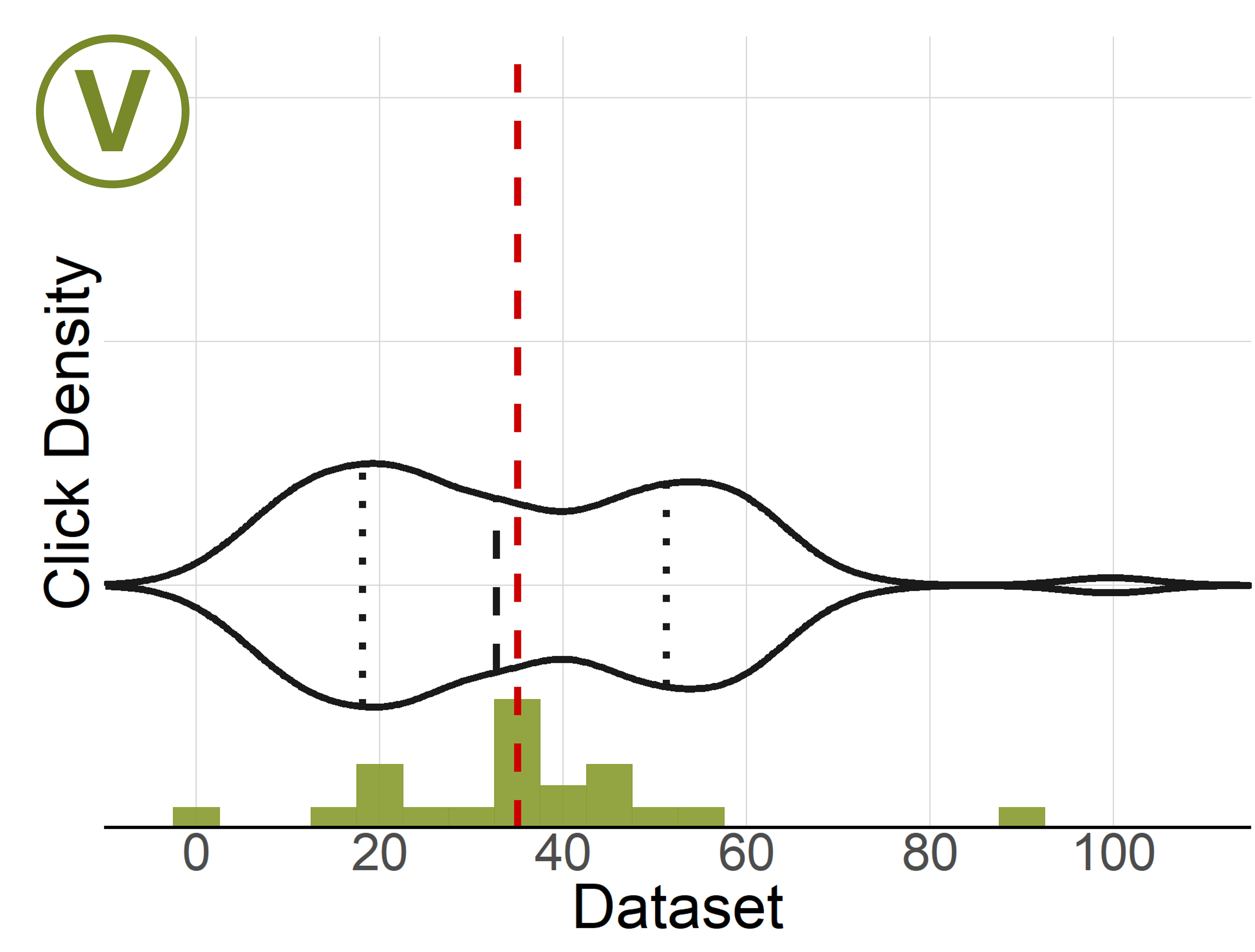}
    \end{subfigure}
        ~ 
    \begin{subfigure}[t]{0.235\textwidth}
        \centering
        \includegraphics[width=.99\textwidth]{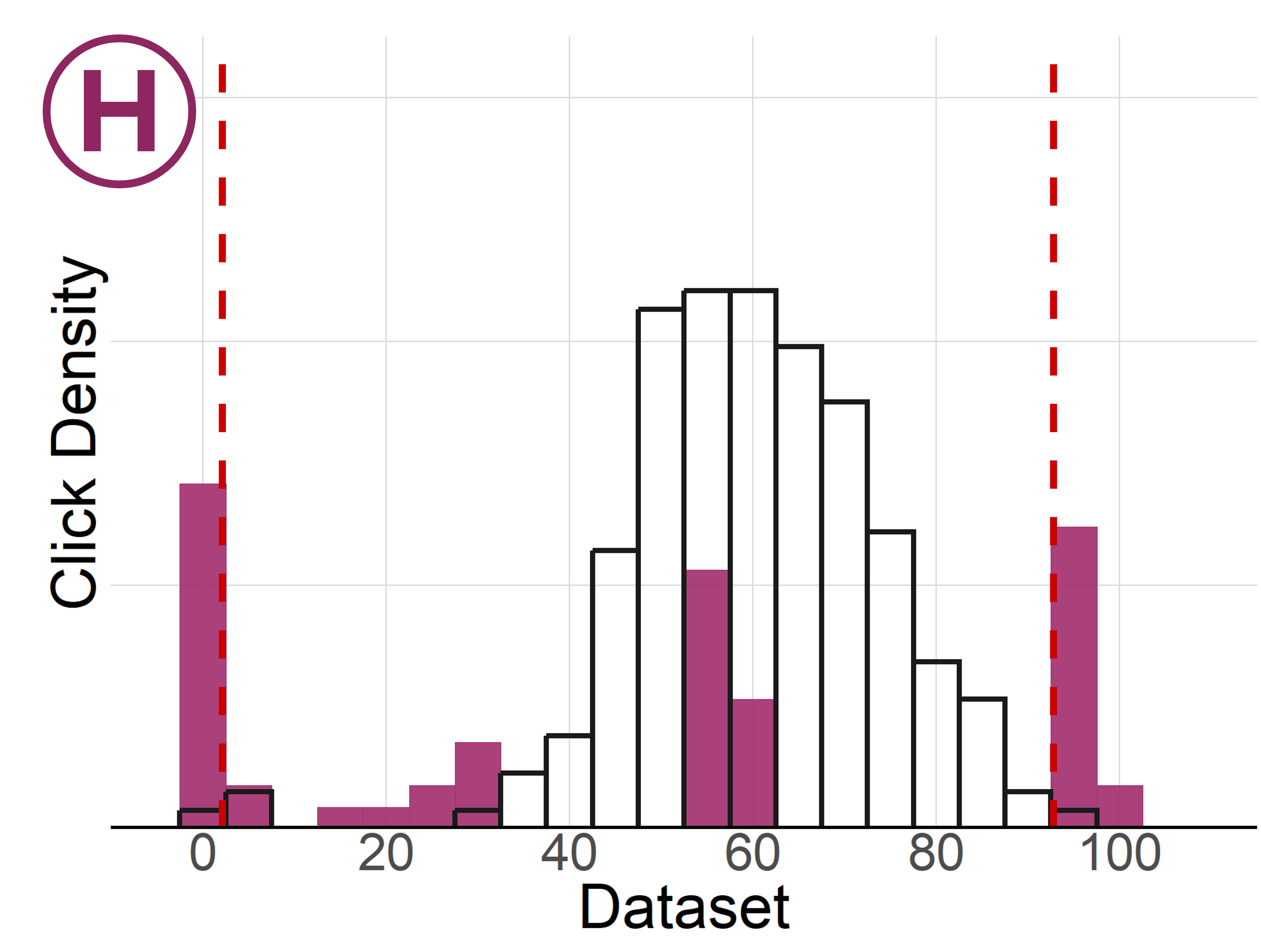}
    \end{subfigure}
        ~
    \begin{subfigure}[t]{0.235\textwidth}
        \centering
        \includegraphics[width=.99\textwidth]{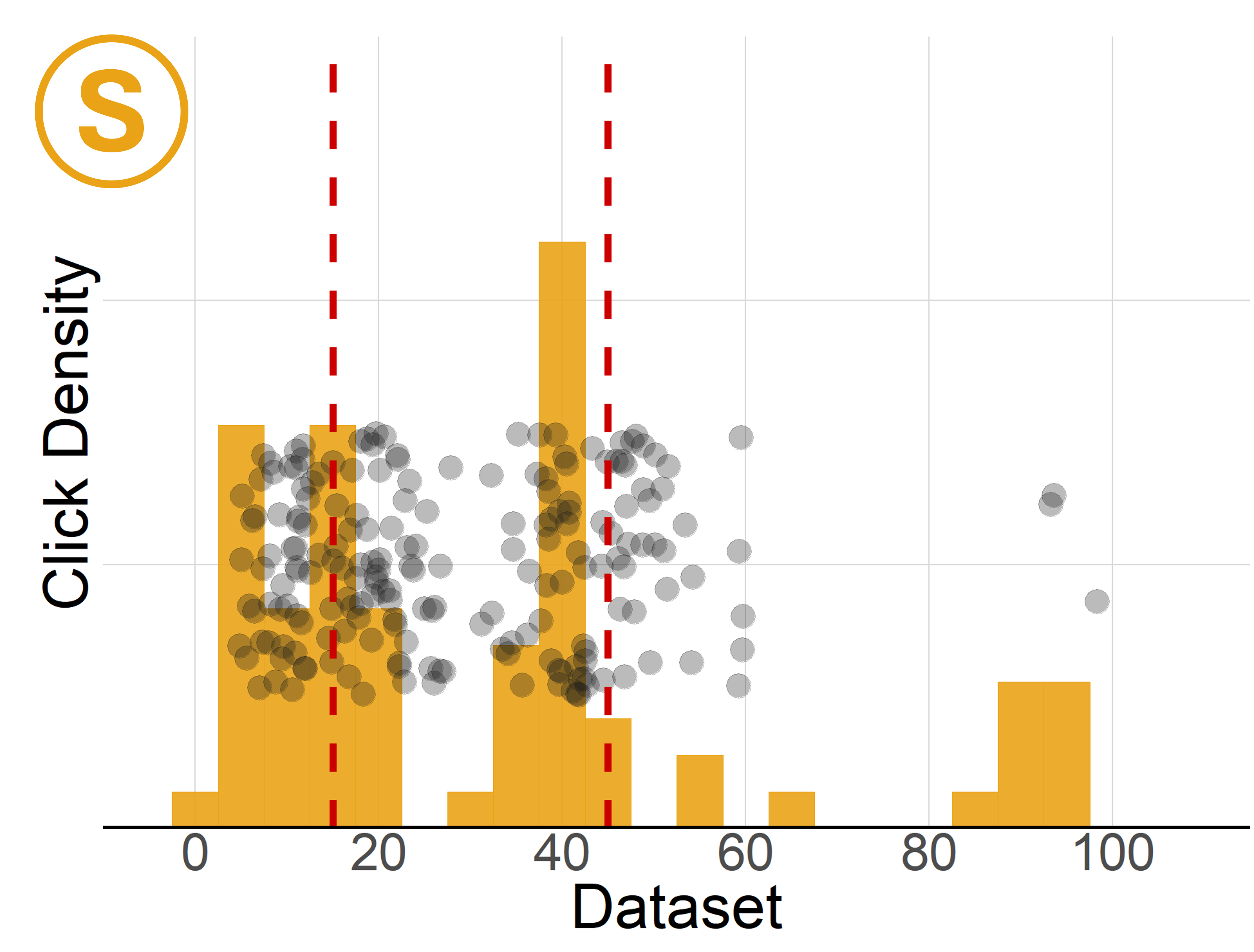}
    \end{subfigure}%
    \caption{Examples of participants' selections. Colored histograms indicate participants' selections, with the outlines of the chart the selections refer to in black and the correct value in red. For better readability, only the coordinates of valid clicks (i.e., clicks on the chart) along the data axis are displayed. \emph{\Boxplot} \Range: Participants mistake both the range of the box, and the end of the whiskers for the range of dataset 6. \emph{\Violinplot} \Mean: Clicks are spread out across large parts of dataset 6. \emph{\Histogram} \Range: The highest bar is falsely identified as the largest value of dataset 1. \emph{\Stripplot} \Clusters: Outliers are falsely identified as an additional cluster in dataset 8. }\label{fig:common_mistakes}
\end{figure*}

\paragraph*{Points of Interest.}
Each chart exhibits distinct patterns for selected points of interest. Figure~\ref{fig:charts_clickmap} illustrates this on one distinct example per chart.
Boxplots (Figure~\ref{fig:charts_clickmap}~\Boxplot) show a large concentration of interest points at the designated median line. Plots that show explicitly encoded outliers also have multiple clicks in their approximate location. Participants also tended to mark at the 25\% and 75\% quartiles, notably often at the corners of the box, and the whiskers were marked at the exact endpoints.
The violinplot (Figure~\ref{fig:charts_clickmap}~\Violinplot) also shows a distinct pattern along the median line, as well as similar markings along the 25\% and 75\% quartiles. Both peaks and pronounced valleys of distribution curves were frequently selected as points of interest for violinplots and for histograms (Figure~\ref{fig:charts_clickmap}~\Histogram), which were marked predominantly at the height of the individual bars. Notably, points where distributions overlap are not marked as prominently as in other charts. 
Lastly, with fewer pronounced geometric features, jittered stripplots (Figure~\ref{fig:charts_clickmap}~\Stripplot) show a wider distribution of interest points than other charts. Selections form clusters in denser areas of the plot and, on fewer occasions, in particularly sparse areas. While modes of bimodal distributions are not selected as accurately as in other charts, we see a concentration of clicks towards the approximate locations of distribution centers.

\paragraph*{Insights.}
The 1633 insights divide up roughly evenly across the charts, with \Boxplot~boxplot (407) and \Violinplot~violinplot (439) being assigned slightly more than \Histogram~histogram (404) and jittered \Stripplot~stripplot (383).
However, a disproportionately high number of participants reported the explicitly encoded \textbf{medians} (210) in the boxplot (53\%) and violinplot (36\%).
The general \textbf{shape} (228) of the distribution was often discussed by participants (e.g., \textit{The data form a clear bell-shaped curve centered around 60}). 
When participants reasoned about \textbf{modes} (119) in the data, this occurred most frequently for the histogram (55\%), more often than for the jittered stripplot (18\%) and the violinplot (25\%). 
\textbf{Clusters} (282) (e.g., \textit{high concentration around 15-25}) were reported predominantly for jittered stripplots (47\%) and equally often for histogram and violinplots (27\%). Notably, the boxplot (42\%) and the violinplot (30\%) motivated participants to reason about the \textbf{symmetry} (64) of the distribution. 
Among comments that refer to the \textbf{range} (123) of data in the plot, we observe that more participants remarked on this in conjunction with the boxplot (42\%).
\textbf{Outliers} (360) were pointed out relatively often during free exploration, especially for jittered stripplots (29\%) while notably fewer participants did so for the violinplot (20\%) than for others.
Finally, we counted a number of obvious \textbf{misinterpretations} (102). 
The highest number of misinterpretations (62\%) occurred in the boxplot. We found several statements (32) misinterpreting the box as a cluster of values, and overall, several difficulties with reading the representation.
Across all charts, roughly equally often (7-9 times each), we observed people confusing univariate distributions with time-dependent charts.

\paragraph*{Common Mistakes. }
In each of the benchmark tasks, participants systematically made the same mistakes. Figure~\ref{fig:common_mistakes} illustrates this on one distinct example per chart.
For \Range, we found that participants mistook the lowest and highest bar in histograms, the widest area of the violinplot, and the limits of the box in boxplots for the range and did not include outlier values in the range of boxplots and histograms (Figures~\ref{fig:common_mistakes}~\Boxplot~and~\Histogram). Violinplots not showing the exact range due to the smoothing of the curve is also reflected in the results. Both \Mean~and \Median~showed more distributed clicks in both bimodal and skewed distributions in all charts, with many clicks on the explicit median encoding where present (Figure~\ref{fig:common_mistakes}~\Violinplot). A common pattern we observed in identifying \Clusters~in histograms and jittered stripplots was distinguishing them from outliers (Figure~\ref{fig:common_mistakes}~\Stripplot), an issue less present with violinplots. The majority of participants identified the box of a boxplot as a cluster rather than recognizing that this information is not available in a boxplot. 

\paragraph*{Open Feedback.}
The \textbf{explicit encoding} of summary statistics was the most frequently mentioned helpful factor for boxplots and violinplots. 
However, the high degree of \textbf{abstraction} in boxplots was considered a limiting factor (\textit{it hides the distribution shape}). 
Conversely, the presence of \textbf{raw data} points helped some participants interpret jittered stripplots, which they described as more transparent (\textit{offering a broader view of the raw data}).
In terms of general usability, histograms and violinplots were often identified as \textbf{intuitive} for helping participants identify interesting points such as peaks, while participants notably struggled in interpreting boxplots, often having trouble interpreting them (\textit{it confuses me a bit}).
Compared with all other charts, histograms were also more often referred to as \textbf{familiar} by participants.
While many participants found histograms clear in terms of visual quality (\textit{Histogram shows how the values are spread out clearly.}), jittered stripplots were more often scrutinized in this regard, especially due to the cluttering of data points (\textit{too cluttered and can be difficult to read the data}) and low potential scalability.

\subsection{Quantitative Results.} We perform visual exploratory analysis on our quantitative results and validate our findings by fitting logistic regression models for binary, and linear mixed-effect models for all other variables, followed by Type-III ANOVA and post-hoc comparisons, with multiple-comparison corrections applied via Benjamini–Hochberg for binary variables and Tukey adjustments for all non-binary variables. For analyses spanning all tasks or involving ratings, we instead fit linear mixed-effects models to account for repeated measurements per participant or participant-specific tendencies. We report differences between charts as \textit{significant}, when the p-value is below $<0.05$. A table of exact p-values and model details, along with figures containing all charts and tasks, can be found in the supplemental material.

\paragraph*{Influence of Participant Background.} While we did find differences in performance between participants with different levels of education and visualization/statistics knowledge, the overall patterns in the charts identified visually and discussed below largely hold across these groups. To account for these performance differences driven by participants' professional backgrounds~\cite{hall_professional_2022} and to make sure the results we report hold across participant groups, we control for education, field of study, chart familiarity, and domain background in all statistical models.

\begin{figure}[!t]
    \centering
    \includegraphics[width=\linewidth]{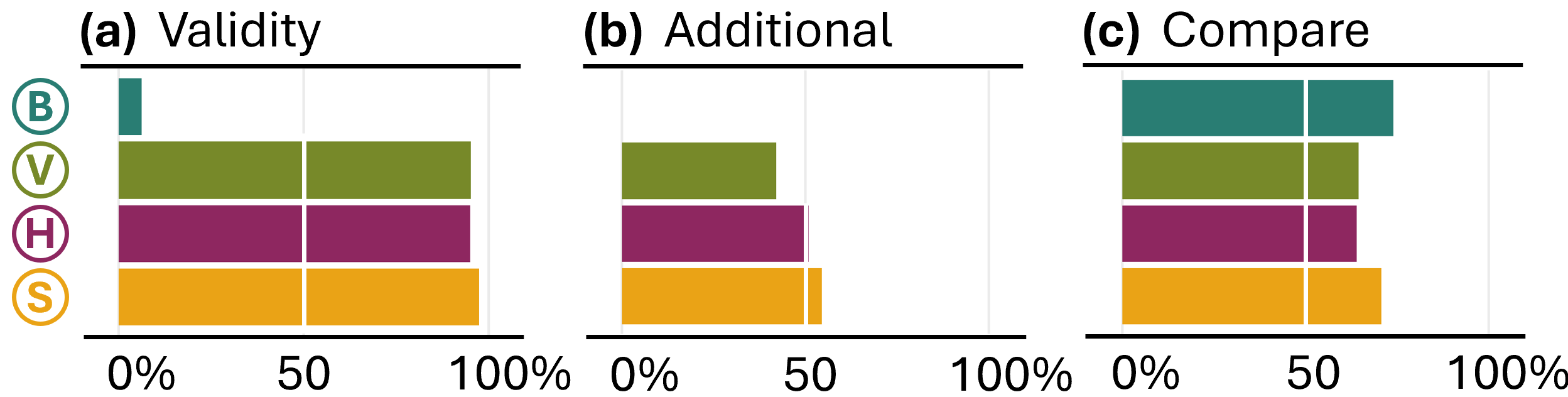}
    \vspace{-3mm}
    \caption{Percentage of \textbf{(a)} valid clicks for the \Clusters~task, \textbf{(b)} additional clicks for the \Clusters~task, and \textbf{(c)} correct comparisons for both \Median~and \Mean. Because no correct value exists for the clusters task in boxplots, all clicks on the chart are counted as \emph{invalid}, so no additional clicks exist by definition.}
    \label{fig:binary}
\end{figure}

\paragraph*{Binary Task Performance Variables.}
We examined four binary task outcomes: the validity of each click, the correctness of the comparison task, if clicks were missing, and the type of each click (closest or additional). 
For \Clusters, our analysis shows a significantly higher percentage of invalid clicks for boxplots across all levels of education (Figure~\ref{fig:binary}a), and a significantly higher percentage of additional clicks in jittered stripplots and histograms than in violinplots (Figure~\ref{fig:binary}b), indicating a clearer distinction between clusters and outliers in violinplots than the other two charts. Across the two comparison tasks, our results show a significantly higher percentage of correct comparisons in boxplots than in histograms or violinplot (Figure~\ref{fig:binary}c). 

\begin{figure}[t!]
    \centering
    \includegraphics[width=\linewidth]{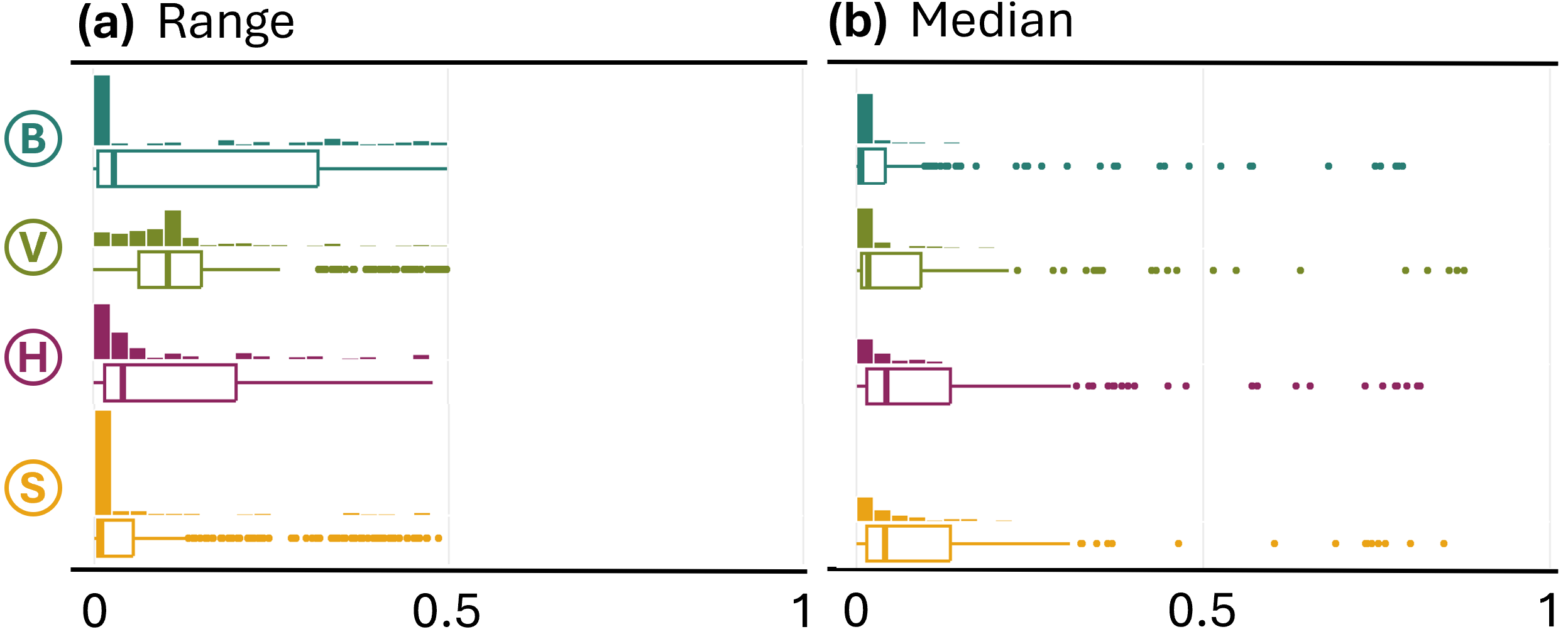}
    \caption{Distributions of relative errors for \textbf{(a)} the \Range~and \textbf{(b)} the \Median~task that clearly show the superior performance of jittered stripplots and inferior performance of boxplots for identifying the range, and the superior performance of charts with explicit encodings for identifying the median.}
    \label{fig:accuracy}
\end{figure}

\paragraph*{Task Accuracy.} For all clicks classified as \textit{valid} and \textit{closest}, we further evaluate the accuracy (i.e., relative error). Across all tasks, our analysis reveals a significant difference in accuracy between boxplots and jittered stripplots, with a lower median, but wider interquartile range and accumulation of responses away from zero error in boxplots, indicating systematic misinterpretation of the chart rather than unsystematic estimation noise.
This result holds for \Range, for which we see a significantly higher accuracy for jittered stripplots and a lower accuracy for boxplots, see Figure~\ref{fig:accuracy}a. Again, the larger interquartile range for both boxplots and histograms we see in the figure indicates a higher number of participants entirely missing the range in these two charts, as opposed to inaccurate estimations in violinplots. 
In contrast, for \mbox{\Median,} our results reveal a significantly higher accuracy in boxplots than in histograms and jittered stripplots, see Figure~\ref{fig:accuracy}b. Overall, the accuracy for identifying \Clusters~is similar across chart types (except boxplots), but violinplots are significantly more accurate for identifying the lower mode of a bimodal distribution, whereas jittered stripplots show the widest spread in relative errors for single-cluster identification, indicating that participants systematically selected areas as clusters that they did not mistake for clusters in the other charts.

\begin{figure}[b!] 
\vspace{-2mm}
    \centering
    \includegraphics[width=\linewidth]{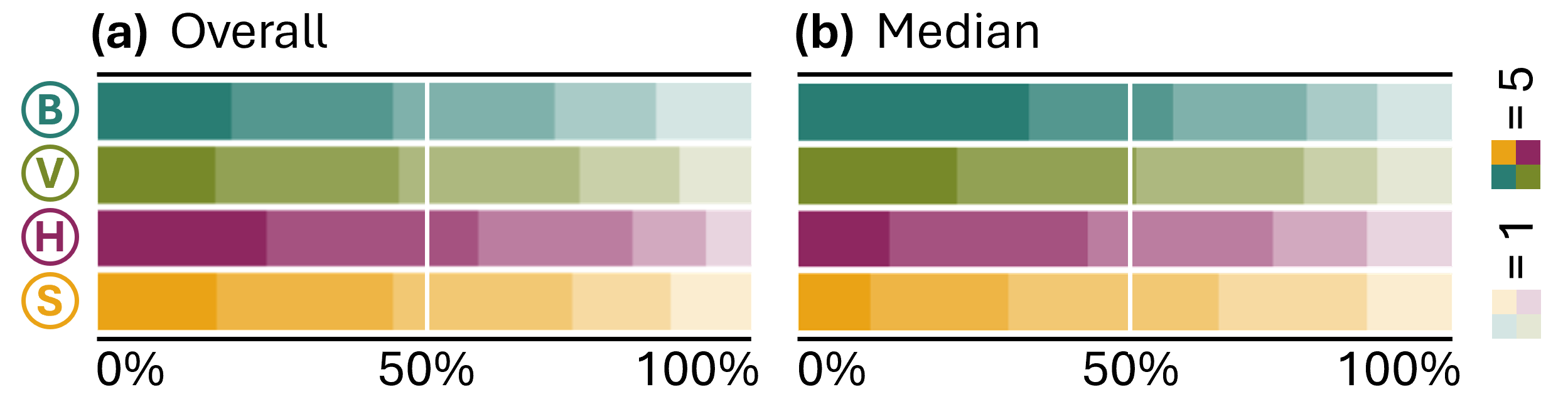}
    \vspace{-3mm}
    \caption{Confidence ratings \textbf{(a)} across all tasks and \textbf{(b)} for the \Median~task. Overall, participants were most confident with histograms, but the explicit encoding of the median leads to a higher confidence in boxplots and violinplots. }
    \label{fig:confdiff}
\end{figure}

\paragraph*{Confidence \& Perceived Difficulty.} Across all tasks, we find a significant difference in both difficulty and confidence between histograms
and all other charts, see Figure~\ref{fig:confdiff}a.
In particular, identifying the \Range~and identifying \Clusters~were rated significantly less difficult, and with higher confidence, for histograms than for the other charts, with lower confidence for \Clusters~in boxplots than in other charts.
Conversely, identifying the \Median~was rated significantly less difficult, and with higher confidence, in charts with an \textbf{explicit encoding}, see Figure~\ref{fig:confdiff}b. 

\paragraph*{Chart Preference.}
Our analysis reveals a significant overall %($p_{corr} < 0.001$) 
preference for histograms over all other charts both during the survey and for future use. This also holds when combining individual task preferences across tasks, and for all individual tasks except identifying the \Median, for which the boxplot (likely due to its \textbf{explicit encoding}) ranked significantly higher than the other charts, with jittered stripplots the least preferred.
Additionally, we find that for \Range, jittered stripplots are significantly preferred over boxplots, and for \Clusters, boxplots are significantly the least preferred charts. 

\vspace{-2mm}

\section{Interviews with Domain Experts}
We conducted semi-structured interviews to determine how domain experts read and design visual representations of univariate data distributions, with five experts from different domains: \textbf{Stat} (PhD, statistics) has been working as an applied data science consultant for four years; \textbf{Vis} (PhD, data visualization) has eight years of research experience; \textbf{Med} (MD, PhD candidate in medicine) has ten years of research experience; \textbf{Env1} and \textbf{Env2} (PhDs, environmental chemistry) have five and seven years of research experience, respectively.
The experts were asked which charts they encounter and use, for which tasks, and for which data. Interviews lasted 30–60 minutes and were conducted via Zoom.

\paragraph*{Tasks and Data.} All domain experts working in academia (Vis, Med, Env1, Env2) mentioned \textbf{comparison} of several, often more than two, groups as one of their main tasks. Usually, they compare summary statistics, such as the mean and standard deviation, or the median and quartiles. All five experts create distribution charts both to \textbf{report findings} (especially to communicate significant differences between groups) and for their own \textbf{exploratory analysis} (to understand the general structure of higher-dimensional data (Stat), to check the results of their classification algorithms (Env2), or to understand the distribution to decide on further analysis and charts (Med)). Sample sizes are highly variable, even within a field, ranging from three data points in certain environmental chemistry studies to more than a million in others. 

\paragraph*{Chart Preferences.} Charts are selected by trying out different visualizations to find one that shows the most information (Env2, Vis), depending on dataset size (Env1), data distribution (Med), and audience (Stat, Med), or by simply defaulting to standard charts without too much deliberation (all participants). For example, Med mentioned that \textit{``medical researchers need explicit encodings of summary statistics''} and that for them \textit{``a boxplot contains all information [..] needed''}. 
To present results, all experts commonly use boxplots, bar charts with error bars, and violinplots. Despite initial confusion about a boxplot's content, including references to means, standard deviations, and confidence intervals, all experts demonstrated correct understanding when asked directly. Histograms are preferably used for data exploration, both because their bars are easily misunderstood by certain audiences and because they do not allow easy comparison of many groups. The jittered stripplot is commonly used by Env1, who usually works with small datasets, but is not appreciated by the other experts due to its meaningless width and overplotting. However, Stat mentioned that they use it to highlight unclear results. 

\paragraph*{Insights.} All experts appreciated the interview as a reminder to be more deliberate in their chart choices rather than defaulting to standard ones. In particular, Med and Env1 expressed intent to increasingly use violinplots in their future work. Notably, Stat, who stated an extensive use of different charts and is experienced with data from a wide range of domains, only realized upon seeing our charts how much information gets lost in a boxplot, stating \textit{``I just had the interesting realization that bimodality is completely lost in a boxplot''}.

\section{Discussion}
Before discussing the implications of our results and how they generalize to other charts, we briefly relate them to our research questions.

\subsection{RQ-1: Performance and Experience} 
Our results show that the abstract features of \Boxplot~boxplots, such as the box, are prone to misinterpretation but useful for tasks related to the explicitly encoded summary statistics, as is also found in \Violinplot~violinplots. Their smooth distribution curve makes violinplots particularly useful for tasks related to the distribution shape, while it obscures the exact range and leads to frequent confusion with time-dependent charts.  \Histogram~Histograms perform worse than other charts for the comparison of median and mean, and the vertical axis is frequently misunderstood, but participants generally feel confident working with them. Jittered \Stripplot~stripplots are optimal for identifying the range of a dataset and helpful for comparing the mean and median of two datasets, but hinder the identification of areas of interest. Overall, we find that \textbf{performance differs systematically by chart type and task} and that \textbf{common misunderstandings are chart-specific,} with \textbf{explicit encodings supporting the retrieval of summary statistics}.

\subsection{RQ-2: Practices and Preferences} 
\Boxplot~Boxplots are preferred mainly for identifying the median and other explicitly encoded summary statistics, and otherwise considered too abstract, non-intuitive, or showing too little detail, however, they are heavily used across research domains, especially for comparison tasks. Participants and domain experts were least familiar with \Violinplot~violinplots, but generally found them intuitive. \Histogram~Histograms were clearly preferred by survey participants, in part due to their high familiarity with them, but are not commonly used across research domains. Jittered \Stripplot~stripplots were found too cluttered and non-intuitive, but appreciated for their transparency and usefulness for small datasets. Overall, we find that \textbf{familiarity influences preferences} and, similarly, \textbf{expert practice is mostly shaped by convention.}

\subsection{Key Findings and Generalization} 
Combining qualitative and quantitative insights, one of our main findings is that \textbf{performance and preference of charts do not always align}. To further underline this, we condense the results into a three-point rating (\good~good, \neutral~neutral, \bad~bad) per research question for each of our benchmark tasks (Table~\ref{tab:results}). The \emph{comparison} of \Mean~and \Median~is listed separately, because our results indicate differences both in performance and preference between identifying and comparing these values.
Below, we discuss our findings for each of the charts we used as stimuli and how they might (not) generalize to similar charts.

\begin{table}[h!]
\captionsetup{justification=justified,singlelinecheck=false}
\caption{Overview of results (\good~good, \neutral~neutral, \bad~bad) regarding RQ-1 (left) and RQ-2 (right). }
\label{tab:results}
\centering
\begin{tabular}{lcccc} & \Boxplot  & \Violinplot  & \Histogram & \Stripplot  \\ 
\vspace{-3mm}\\
\toprule 
\Range     &   \bad~\bad       &     \bad~\bad        &   \neutral~\good  &   \good~\good            \\
\Mean~{\sffamily\textsc{(find)}}     &   \neutral~\bad   &     \neutral~\neutral     &   \neutral~\good  &   \neutral~\neutral         \\
\Median~{\sffamily\textsc{(find)}}     &   \good~\good     &     \good~\good       &   \bad~\bad       &   \neutral~\neutral            \\
{\sffamily\textsc{Comparison}}    &  \good~\good     &  \neutral~\good       &   \bad~\neutral   &   \good~\bad    \\
\Clusters  &   \bad~\bad       &     \good~\good        &   \good ~\good  &   \neutral~\neutral   \\
\bottomrule
\end{tabular}
\end{table}

%Boxplots difficult
\noindent \includegraphics[height=1em]{B.png} Our results confirm and extend previously~\cite{lem_experts_2014, lem_misinterpretation_2013} identified common misinterpretations of \Boxplot~boxplots, in particular of the box. While these misinterpretations could potentially be avoided using box-less charts displaying summary statistics~\cite{tufte_envisioning_1990, carr_colorful_1994}, the high level of abstraction, missing details, and lack of intuitiveness criticized by participants in this and other~\cite{blumenschein_v-plots_2020} studies remain. 
Despite these limitations well known to the domain experts we interviewed, the interviews also revealed a heavy use of boxplots across domains, mainly due to their familiarity with scientific audiences, which makes it difficult to argue against their use.

\noindent \includegraphics[height=1em]{V.png}  
Smooth densities facilitate tasks related to the distribution shape, such as identifying clusters and characterizing the distribution's symmetry, which provides performance-based evidence for the preference-based findings by Blumenschein et al.~\cite{blumenschein_v-plots_2020}. Especially in situations where a distinction between outlying values and data clusters is more important than the identification of exact values, \Violinplot~violinplots are superior to both histograms and jittered stripplots. These results are directly related to the continuous density curve, shown to be violinplots' main strength~\cite{molina_how_2022}, and should thus generalize to other such charts. Asymmetric charts, however, might be more prone to misunderstandings related to a suspected time-dependent nature of the data that we see with violinplots~\cite{correll_error_2014}. 
While all domain experts we interviewed stated that violinplots are not common in their field, and participants were least familiar with them, both groups found them generally intuitive and powerful, highlighting the potential in better integrating them into science practices in various research domains~\cite{debbeler_polarized_2018}.

\noindent \includegraphics[height=1em]{H.png} 
For \Histogram~histograms, we observed a preference pattern opposite to that of boxplots: survey participants were highly familiar and showed a clear preference, whereas our domain experts do not use them frequently due to small sample sizes, the need to compare several groups simultaneously, and difficulties in explaining the vertical axis of histograms to their audiences. 
While, similar to previous studies~\cite{lem_experts_2014, lem_misinterpretation_2013}, the most common errors our participants made with histograms appear to stem from a misinterpretation of the vertical axis, our quantitative findings offer little support for this concern: histograms did not perform significantly worse than the other charts, except for summary-statistic comparisons, indicating that domain experts should base their decisions on data and tasks rather than assumptions about their audiences.
Representations that encode binned densities in different visual variables than bar height, for example, using color~\cite{heim_accustripes_2024} or dots representing a certain number of data points~\cite{wilkinson_dot_1999}, might help mitigate such effects. 

\noindent \includegraphics[height=1em]{S.png} The significantly higher percentage of additional clicks in \Clusters~and the wide distribution of interest points in \Describe~ indicate that jittered \Stripplot~stripplots hinder pinpointing relevant parts of a dataset without a shape to guide the eye~\cite{molina_how_2022}. This effect is confirmed by survey participants considering them too cluttered, and domain experts finding them non-intuitive. Even the high performance in identifying the \Range~of the dataset is contrasted with a lower confidence and greater perceived difficulty than in histograms. Still, both groups appreciated them for their transparency and usefulness for small datasets, so arranging the data points to avoid overplotting and show the distribution shape, such as dotplots~\cite{wilkinson_dot_1999} and its variants~\cite{rodrigues_nonlinear_2018, rodrigues_relaxed_2023} or beeswarm plots~\cite{eklund_beeswarm_2010} may retain these advantages while avoiding drawbacks.
Our surprising results that jittered stripplots outperform violinplots in comparing \Mean~and \Median~of two datasets might indicate that dense point clouds convey central tendency more intuitively than abstract smooth curves, and aligns with findings related to comparing means in scatterplots~\cite{gleicher_perception_2013}. 

Rather than identifying a universally superior distribution chart, our results point to a relationship between user performance and preference and the combination of chart, task, and audience. This insight reinforces the \textbf{need for task-aware and audience-aware design decisions}, rather than reliance on convention or familiarity. The systematic trade-offs observed across charts further provide \textbf{empirical motivation for defensive and hybrid approaches}~\cite{correll_teru_2023}, such as v-plots~\cite{blumenschein_v-plots_2020} and raincloud plots~\cite{allen_raincloud_2021, correll_teru_2023}, to mitigate the weaknesses of any single representation by combining complementary encodings. While our conclusions are based on specific charts implemented with common \texttt{ggplot} defaults, not all patterns may generalize beyond these charts~\cite{skau_evaluation_2015}, our study offers an empirical baseline for widely used defaults.  
In contrast to prior work using binary \textit{true/false} metrics~\cite{lem_experts_2014, lem_misinterpretation_2013, rodrigues_comparing_2019}, our \textbf{click-to-select approach captures \textit{how} people read and (mis)understand charts}. The clear, systematic click patterns indicate that many participants understood the tasks and completed them with care rather than clicking at random. In combination with a substantial proportion of participants providing detailed responses to the free-text question, this approach enabled us to gain a deeper understanding of participants' reasoning about the charts.

\subsection{Limitations \& Future Work}
While many of our findings on participant perception and frequent errors in charts are related to key characteristics of the four chart groups described in Section~\ref{subsec:related_charts}, minor design changes even within the same chart type may impact both preferences and performance~\cite{correll_looks_2019}, and further research is thus needed to generalize our results beyond the \texttt{ggplot} default plots used in this study. 
Likewise, variations of our benchmark tasks, or tasks that more closely resemble real-world analytical situations, could help validate and refine our insights in more realistic analytical contexts.
Our click-to-select approach is promising for future research on chart perception beyond quantitative performance metrics, and could, for example, extend prior work~\cite{correll_teru_2023} on hybrid charts, such as the v-plot~\cite{blumenschein_v-plots_2020} and the raincloud plot~\cite{allen_raincloud_2021, correll_teru_2023}.
Finally, the results of our interviews with experts across domains indicate that visualization preferences and practices can vary substantially, suggesting that domain-specific studies may provide a richer understanding of how chart choice aligns with disciplinary needs and conventions.

%-------------------------------------------------------------------------
\section{Conclusion}
In this work, we study four distinct groups of visualizations for univariate data distributions in terms of performance and participant experience.
A mixed-methods study based on a click-to-select approach, combining quantitative performance measures and preference ratings with qualitative data on how participants interacted with the charts and expert interviews, provides a comprehensive perspective and performance-based evidence for previous results.
Our results indicate that chart preference and familiarity do not necessarily align with participants’ task performance, and that charts widely liked by general audiences (such as histograms) or commonly used in scientific domains (such as boxplots) are not inherently the most effective for all tasks.
Tailored solutions such as violinplots are received positively by domain experts and study participants alike, and jittered stripplots, which visually overwhelmed many participants, yielded unexpectedly high performance.
In general, chart preferences appear highly field-dependent and shaped more by convention than by analytic suitability.
Consequently, we recommend supporting the interpretability of familiar but potentially confusing concepts through explicit encodings and mitigating confusion by abstraction through the inclusion of raw data points. 

%------------------------------------------------------------------------

\section*{Acknowledgements}
We used ChatGPT solely for grammar correction and language polishing.

% bibtex
\bibliographystyle{alpha} 
\bibliography{egbibsample}       

@book{tukey_exploratory_1977,
    address = {Reading, Mass.},
    series = {Addison-{Wesley} series in behavioral sciences},
    title = {Exploratory data analysis},
    isbn = {0-201-07616-0 978-0-201-07616-5},
    language = {English},
    publisher = {Addison-Wesley Pub. Co.},
    author = {Tukey, John W.},
    year = {1977},
}

@inproceedings{amar_low-level_2005,
    title = {Low-level components of analytic activity in information visualization},
    doi = {10.1109/INFVIS.2005.1532136},
    booktitle = {{IEEE} {Symposium} on {Information} {Visualization}, 2005.},
    author = {Amar, R. and Eagan, J. and Stasko, J.},
    year = {2005},
    pages = {111--117},
}

@book{bloom_taxonomy_1956,
    address = {New York},
    title = {Taxonomy of {Educational} {Objectives}, {Handbook}: {The} {Cognitive} {Domain}},
    urldate = {2025-10-02},
    publisher = {David McKay},
    author = {Bloom, Benjamin},
    year = {1956},
}

@inproceedings{rodrigues_comparing_2019,
    title = {Comparing the {Effectiveness} of {Visualizations} of {Different} {Data} {Distributions}},
    doi = {10.1109/SIBGRAPI.2019.00020},
    booktitle = {2019 32nd {SIBGRAPI} {Conference} on {Graphics}, {Patterns} and {Images} ({SIBGRAPI})},
    author = {Rodrigues, Ariane M. B. and Barbosa, Gabriel D. J. and Lopes, Hélio and Barbosa, Simone D. J.},
    year = {2019},
    pages = {84--91},
}

@article{blumenschein_v-plots_2020,
    title = {v-plots: {Designing} {Hybrid} {Charts} for the {Comparative} {Analysis} of {Data} {Distributions}},
    volume = {39},
    doi = {10.1111/cgf.14002},
    language = {en},
    number = {3},
    journal = {Computer Graphics Forum},
    author = {Blumenschein, Michael and Debbeler, Luka J. and Lages, Nadine C. and Renner, Britta and Keim, Daniel A. and El-Assady, Mennatallah},
    year = {2020},
    pages = {565--577},
}

@article{correll_error_2014,
    title = {Error {Bars} {Considered} {Harmful}: {Exploring} {Alternate} {Encodings} for {Mean} and {Error}},
    volume = {20},
    doi = {10.1109/TVCG.2014.2346298},
    language = {en},
    number = {12},
    journal = {IEEE Transactions on Visualization and Computer Graphics},
    author = {Correll, Michael and Gleicher, Michael},
    year = {2014},
    pages = {2142--2151},
}

@article{hofmann_letter-value_2017,
    title = {Letter-{Value} {Plots}: {Boxplots} for {Large} {Data}},
    volume = {26},
    issn = {1061-8600},
    shorttitle = {Letter-{Value} {Plots}},
    doi = {10.1080/10618600.2017.1305277},
    number = {3},
    journal = {Journal of Computational and Graphical Statistics},
    author = {Hofmann, Heike and Wickham, Hadley and Kafadar, Karen},
    year = {2017},
    pages = {469--477},
}

@article{mcgill_variations_1978,
    title = {Variations of {Box} {Plots}},
    volume = {32},
    issn = {0003-1305},
    doi = {10.2307/2683468},
    number = {1},
    urldate = {2025-09-22},
    journal = {The American Statistician},
    author = {McGill, Robert and Tukey, John W. and Larsen, Wayne A.},
    year = {1978},
    pages = {12--16},
}

@article{hubert_adjusted_2008,
    title = {An adjusted boxplot for skewed distributions},
    volume = {52},
    doi = {10.1016/j.csda.2007.11.008},
    language = {en},
    number = {12},
    journal = {Computational Statistics \& Data Analysis},
    author = {Hubert, M. and Vandervieren, E.},
    year = {2008},
    pages = {5186--5201},
}

@article{benjamini_opening_1988,
    title = {Opening the {Box} of a {Boxplot}},
    volume = {42},
    doi = {10.1080/00031305.1988.10475580},
    number = {4},
    urldate = {2025-09-29},
    journal = {The American Statistician},
    author = {Benjamini, Yoav},
    year = {1988},
    pages = {257--262},
}

@article{north_toward_2006,
    title = {Toward measuring visualization insight},
    volume = {26},
    doi = {10.1109/MCG.2006.70},
    number = {3},
    urldate = {2025-10-28},
    journal = {IEEE Computer Graphics and Applications},
    author = {North, Chris},
    year = {2006},
    pages = {6--9},
}

@book{spear_charting_1952,
    title = {Charting {Statistics}},
    publisher = {McGraw-Hill Book Company, Inc},
    author = {Spear, Mary E.},
    year = {1952},
}

@article{haemer_range-bar_1948,
    title = {Range-{Bar} {Charts}},
    volume = {2},
    doi = {10.1080/00031305.1948.10501576},
    language = {en},
    number = {2},
    urldate = {2025-10-30},
    journal = {The American Statistician},
    author = {Haemer, Kenneth W.},
    year = {1948},
    pages = {23--23},
}

@Manual{R_Core_2021,
     title = {R: A Language and Environment for Statistical Computing},
     author = {{R Core Team}},
     organization = {R Foundation for Statistical Computing},
     address = {Vienna, Austria},
     year = {2021},
     url = {https://www.R-project.org/},
   }

@article{kay_ggdist_2024,
    title = {ggdist: {Visualizations} of {Distributions} and {Uncertainty} in the {Grammar} of {Graphics}},
    volume = {30},
    doi = {10.1109/TVCG.2023.3327195},
    number = {1},
    journal = {IEEE Transactions on Visualization and Computer Graphics},
    author = {Kay, Matthew},
    year = {2024},
    pages = {414--424},
}

@inproceedings{kay_when_2016,
    title = {When (ish) is {My} {Bus}?: {User}-centered {Visualizations} of {Uncertainty} in {Everyday}, {Mobile} {Predictive} {Systems}},
    doi = {10.1145/2858036.2858558},
    language = {en},
    booktitle = {Proceedings of the 2016 {CHI} {Conference} on {Human} {Factors} in {Computing} {Systems}},
    publisher = {ACM},
    author = {Kay, Matthew and Kola, Tara and Hullman, Jessica R. and Munson, Sean A.},
    year = {2016},
    pages = {5092--5103},
}

@article{frigge_implementations_1989,
    title = {Some {Implementations} of the {Boxplot}},
    volume = {43},
    doi = {10.2307/2685173},
    number = {1},
    journal = {The American Statistician},
    author = {Frigge, Michael and Hoaglin, David C. and Iglewicz, Boris},
    year = {1989},
    pages = {50--54},
}

@article{carr_colorful_1994,
    title = {A {Colorful} {Variation} {On} {Box} {Plots}},
    journal = {Statistical Computing \& Statistical Graphics Newsletter},
    author = {Carr, Daniel B},
    year = {1994},
}

@book{tufte_envisioning_1990,
    title = {Envisioning {Information}},
    isbn = {978-1-930824-14-0},
    publisher = {Graphics Press},
    author = {Tufte, E.R.},
    year = {1990},
}

@article{allen_raincloud_2021,
    title = {Raincloud plots: a multi-platform tool for robust data visualization},
    doi = {10.12688/wellcomeopenres.15191.2},
    journal = {Wellcome Open Research},
    author = {Allen, Micah and Poggiali, Davide and Whitaker, Kirstie and Marshall, Tom Rhys and van Langen, Jordy and Kievit},
    year = {2021},
}

@article{kampstra_beanplot_2008,
    title = {Beanplot: {A} {Boxplot} {Alternative} for {Visual} {Comparison} of {Distributions}},
    volume = {28},
    doi = {10.18637/jss.v028.c01},
    number = {Code Snippet 1},
    journal = {Journal of Statistical Software},
    author = {Kampstra, Peter},
    year = {2008},
}

@inproceedings{molina_how_2022,
    address = {Oklahoma City, OK, USA},
    title = {How should we design violin plots?},
    doi = {10.1109/VisGuides57787.2022.00006},
    author = {Molina, E. and Viale, L. and Vazquez, P.},
    year = {2022},
    pages = {1--7},
}

@article{hintze_violin_1998,
    title = {Violin {Plots}: {A} {Box} {Plot}-{Density} {Trace} {Synergism}},
    volume = {52},
    doi = {10.1080/00031305.1998.10480559},
    number = {2},
    journal = {The American Statistician},
    author = {Hintze, Jerry L. and Nelson, Ray D.},
    year = {1998},
    pages = {181--184},
}

@article{sidiropoulos_sinaplot_2018,
    title = {{SinaPlot}: {An} {Enhanced} {Chart} for {Simple} and {Truthful} {Representation} of {Single} {Observations} {Over} {Multiple} {Classes}},
    volume = {27},
    doi = {10.1080/10618600.2017.1366914},
    number = {3},
    journal = {Journal of Computational and Graphical Statistics},
    author = {Sidiropoulos, Nikos and Sohi, Sina Hadi and Pedersen, Thomas Lin and Porse, Bo Torben and Winther, Ole and Rapin, Nicolas and Bagger, Frederik Otzen},
    year = {2018},
    pages = {673--676},
}

@article{heim_accustripes_2024,
    title = {{AccuStripes}: {Visual} exploration and comparison of univariate data distributions using color and binning},
    volume = {119},
    doi = {10.1016/j.cag.2024.103906},
    journal = {Computers \& Graphics},
    author = {Heim, Anja and Gall, Alexander and Waldner, Manuela and Gröller, Eduard and Heinzl, Christoph},
    year = {2024},
    pages = {103906},
}

@inproceedings{sahann_histogram_2021,
    title = {Histogram binning revisited with a focus on human perception},
    booktitle = {2021 IEEE Visualization Conference (VIS)},
    doi = {10.1109/VIS49827.2021.9623301},
    author = {Sahann, Raphael and Möller, Torsten and Schmidt, Johanna},
    year = {2021},
    pages = {66--70},
}

@article{lem_misinterpretation_2013,
    title = {On the misinterpretation of histograms and box plots},
    volume = {33},
    doi = {10.1080/01443410.2012.674006},
    number = {2},
    journal = {Educational Psychology},
    author = {Lem, Stephanie and , Patrick, Onghena and , Lieven, Verschaffel and and Van Dooren, Wim},
    year = {2013},
    pages = {155--174},
}

@article{lem_experts_2014,
    title = {Experts’ {Misinterpretation} of {Box} {Plots} – a {Dual} {Processing} {Approach}},
    volume = {54},
    doi = {10.5334/pb.az},
    number = {4},
    journal = {Psychologica Belgica},
    author = {Lem, Stephanie and Onghena, Patrick and Verschaffel, Lieven and Van Dooren, Wim},
    year = {2014},
    pages = {395--405},
}

@article{correll_looks_2019,
    title = {Looks {Good} {To} {Me}: {Visualizations} {As} {Sanity} {Checks}},
    volume = {25},
    doi = {10.1109/TVCG.2018.2864907},
    number = {1},
    journal = {IEEE Transactions on Visualization and Computer Graphics},
    author = {Correll, Michael and Li, Mingwei and Kindlmann, Gordon and Scheidegger, Carlos},
    year = {2019},
    pages = {830--839},
}

@inproceedings{liu_ridgebuilder_2025,
    title = {{RidgeBuilder}: {Interactive} {Authoring} of {Expressive} {Ridgeline} {Plots}},
    doi = {10.1145/3706598.3714209},
    booktitle = {Proceedings of the 2025 {CHI} {Conference} on {Human} {Factors} in {Computing} {Systems}},
    publisher = {ACM},
    author = {Liu, Shuhan and Liu, Yangtian and Li, Junxin and Huang, Yanwei and Shangguan, Yue and Deng, Zikun and Weng, Di and Wu, Yingcai},
    year = {2025},
    pages = {1--18},
}

@article{wilkinson_dot_1999,
    title = {Dot {Plots}},
    volume = {53},
    doi = {10.2307/2686111},
    number = {3},
    journal = {The American Statistician},
    author = {Wilkinson, Leland},
    year = {1999},
    pages = {276--281},
}

@book{wilkinson_grammar_2005,
    address = {New York, NY},
    edition = {Second edition},
    series = {Statistics and {Computing}},
    title = {The grammar of graphics},
    isbn = {978-0-387-24544-7},
    publisher = {Springer},
    author = {Wilkinson, Leland},
    year = {2005},
}

@article{kozlova_visual_2020,
    title = {Visual {Analytics} in {Environmental} {Decision}-{Making}: {A} {Comparison} of {Overlay} {Charts} versus {Simulation} {Decomposition}},
    volume = {4},
    doi = {10.3808/jeil.202000047},
    number = {2},
    journal = {Journal of Environmental Informatics Letters},
    author = {Kozlova, M. and Yeomans, J. S.},
    year = {2020},
}

@article{leisch_neighborhood_2010,
    title = {Neighborhood graphs, stripes and shadow plots for cluster visualization},
    volume = {20},
    doi = {10.1007/s11222-009-9137-8},
    number = {4},
    journal = {Statistics and Computing},
    author = {Leisch, Friedrich},
    year = {2010},
    pages = {457--469},
}

@Book{ggplot,
author = {Hadley Wickham},
    title = {ggplot2: Elegant Graphics for Data Analysis},
    publisher = {Springer-Verlag New York},
    year = {2016},
    isbn = {978-3-319-24277-4},
    url = {https://ggplot2.tidyverse.org},
  }

@article{vanderplas_spatial_2016,
    title = {Spatial {Reasoning} and {Data} {Displays}},
    volume = {22},
    doi = {10.1109/TVCG.2015.2469125},
    number = {1},
    journal = {IEEE Transactions on Visualization and Computer Graphics},
    author = {VanderPlas, Susan and Hofmann, Heike},
    year = {2016},
    pages = {459--468},
}

@misc{Yau2012_VisualizeCompareDistributions,
  author       = {Nathan Yau},
  title        = {How to {Visualize} and {Compare} {Distributions} in {R}},
  year         = {2012},
  day          = {15},
  howpublished = {\url{https://flowingdata.com/2012/05/15/how-to-visualize-and-compare-distributions/}},
  note         = {Accessed: 2025-11-12}
}

@article{shneiderman_eyes_1996,
    title = {The {Eyes} {Have} {It}: {A} {Task} by {Data} {Type} {Taxonomy} for {Information} {Visualizations}},
    journal = {Proceedings of the 1996 IEEE Symposium on Visual Languages},
    author = {Shneiderman, Ben},
    year = {1996},
    pages = {336},
}

@article{douglas_data_2023,
    title = {Data quality in online human-subjects research: {Comparisons} between {MTurk}, {Prolific}, {CloudResearch}, {Qualtrics}, and {SONA}},
    volume = {18},
    doi = {10.1371/journal.pone.0279720},
    number = {3},
    journal = {PLOS ONE},
    author = {Douglas, Benjamin D. and Ewell, Patrick J. and Brauer, Markus},
    year = {2023},
    pages = {e0279720},
}

@article{skau_evaluation_2015,
    title = {An {Evaluation} of the {Impact} of {Visual} {Embellishments} in {Bar} {Charts}},
    volume = {34},
    doi = {10.1111/cgf.12634},
    number = {3},
    journal = {Computer Graphics Forum},
    author = {Skau, Drew and Harrison, Lane and Kosara, Robert},
    year = {2015},
    pages = {221--230},
}

@article{brehmer_multi-level_2013,
    title = {A {Multi}-{Level} {Typology} of {Abstract} {Visualization} {Tasks}},
    volume = {19},
    doi = {10.1109/TVCG.2013.124},
    number = {12},
    journal = {IEEE Transactions on Visualization and Computer Graphics},
    author = {Brehmer, Matthew and Munzner, Tamara},
    year = {2013},
    pages = {2376--2385},
}

@misc{questionpro,
  author       = {{QuestionPro Inc.}},
  title        = {QuestionPro: Online Survey Software},
  year         = {2024},
  howpublished = {\url{https://www.questionpro.com}},
  note         = {Accessed: 2025-11-23}
}

@article{kim_bubbleview_2017,
    title = {{BubbleView}: {An} {Interface} for {Crowdsourcing} {Image} {Importance} {Maps} and {Tracking} {Visual} {Attention}},
    volume = {24},
    shorttitle = {{BubbleView}},
    doi = {10.1145/3131275},
    number = {5},
    urldate = {2025-11-29},
    journal = {ACM Transactions on Computer-Human Interaction},
    author = {Kim, Nam Wook and Bylinskii, Zoya and Borkin, Michelle A. and Gajos, Krzysztof Z. and Oliva, Aude and Durand, Fredo and Pfister, Hanspeter},
    year = {2017},
    pages = {1--40},
}

@article{quadri_survey_2022,
    title = {A {Survey} of {Perception}-{Based} {Visualization} {Studies} by {Task}},
    volume = {28},
    doi = {10.1109/TVCG.2021.3098240},
    number = {12},
    journal = {IEEE Transactions on Visualization and Computer Graphics},
    author = {Quadri, Ghulam Jilani and Rosen, Paul},
    year = {2022},
    pages = {5026--5048},
}

@article{debbeler_polarized_2018,
    title = {Polarized but illusory beliefs about tap and bottled water: {A} product- and consumer-oriented survey and blind tasting experiment},
    volume = {643},
    doi = {10.1016/j.scitotenv.2018.06.190},
    journal = {Science of The Total Environment},
    author = {Debbeler, Luka Johanna and Gamp, Martina and Blumenschein, Michael and Keim, Daniel and Renner, Britta},
    year = {2018},
    pages = {1400--1410},
}

@article{gleicher_perception_2013,
    title = {Perception of {Average} {Value} in {Multiclass} {Scatterplots}},
    volume = {19},
    doi = {10.1109/TVCG.2013.183},
    number = {12},
    journal = {IEEE Transactions on Visualization and Computer Graphics},
    author = {Gleicher, Michael and Correll, Michael and Nothelfer, Christine and Franconeri, Steven},
    year = {2013},
    pages = {2316--2325},
}

@inproceedings{stoiber_visualization_2019,
    title = {Visualization {Onboarding}: {Learning} {How} to {Read} and {Use} {Visualizations}},
    doi = {10.31219/osf.io/c38ab},
    booktitle = {IEEE Workshop on Visualization for Communication},
    publisher = {IEEE},
    author = {Stoiber, Christina and Grassinger, Florian and Pohl, Margit and Stitz, Holger and Streit, Marc and Aigner, Wolfgang},
    year = {2019},
}

@article{correll_teru_2023,
    title = {Teru {Teru} {Bōzu}: {Defensive} {Raincloud} {Plots}},
    volume = {42},
    doi = {10.1111/cgf.14826},
    number = {3},
    journal = {Computer Graphics Forum},
    author = {Correll, Michael},
    year = {2023},
    pages = {235--246},
}

@article{hullman_hypothetical_2015,
    title = {Hypothetical {Outcome} {Plots} {Outperform} {Error} {Bars} and {Violin} {Plots} for {Inferences} about {Reliability} of {Variable} {Ordering}},
    volume = {10},
    doi = {10.1371/journal.pone.0142444},
    number = {11},
    journal = {PLOS ONE},
    author = {Hullman, Jessica and Resnick, Paul and Adar, Eytan},
    year = {2015},
    pages = {e0142444},
}

@article{fygenson_impact_2025,
    title = {Impact of {Vertical} {Scaling} on {Normal} {Probability} {Density} {Function} {Plots}},
    volume = {31},
    doi = {10.1109/TVCG.2024.3456396},
    number = {1},
    journal = {IEEE Transactions on Visualization and Computer Graphics},
    author = {Fygenson, Racquel and Padilla, Lace},
    year = {2025},
    pages = {984--994},
}

@article{rodrigues_nonlinear_2018,
    title = {Nonlinear {Dot} {Plots}},
    volume = {24},
    doi = {10.1109/TVCG.2017.2744018},
    number = {1},
    journal = {IEEE Transactions on Visualization and Computer Graphics},
    author = {Rodrigues, Nils and Weiskopf, Daniel},
    year = {2018},
    pages = {616--625},
}

@article{rodrigues_relaxed_2023,
    title = {Relaxed {Dot} {Plots}: {Faithful} {Visualization} of {Samples} and {Their} {Distribution}},
    volume = {29},
    doi = {10.1109/TVCG.2022.3209429},
    number = {1},
    journal = {IEEE Transactions on Visualization and Computer Graphics},
    author = {Rodrigues, Nils and Schulz, Christoph and Döring, Sören and Baumgartner, Daniel and Krake, Tim and Weiskopf, Daniel},
    year = {2023},
    pages = {278--287},
}

@article{knittel_visual_2021,
    title = {Visual {Neural} {Decomposition} to {Explain} {Multivariate} {Data} {Sets}},
    volume = {27},
    doi = {10.1109/TVCG.2020.3030420},
    number = {2},
    journal = {IEEE Transactions on Visualization and Computer Graphics},
    author = {Knittel, Johannes and Lalama, Andres and Koch, Steffen and Ertl, Thomas},
    year = {2021},
    pages = {1374--1384},
}

@article{stuart_sea_2024,
    title = {Sea stack plots: {Replacing} bar charts with histograms},
    volume = {14},
    doi = {10.1002/ece3.11237},
    number = {4},
    journal = {Ecology and Evolution},
    author = {Stuart, Alice Dorothy and Ilić, Maja and Simmons, Benno I. and Sutherland, William J.},
    year = {2024},
    pages = {e11237},
}

@article{potter_visualizing_2010,
    title = {Visualizing {Summary} {Statistics} and {Uncertainty}},
    volume = {29},
    doi = {10.1111/j.1467-8659.2009.01677.x},
    number = {3},
    journal = {Computer Graphics Forum},
    author = {Potter, K. and Kniss, J. and Riesenfeld, R. and Johnson, C.r.},
    year = {2010},
    pages = {823--832},
}

@misc{eklund_beeswarm_2010,
    title = {beeswarm: {The} {Bee} {Swarm} {Plot}, an {Alternative} to {Stripchart}},
    doi = {10.32614/CRAN.package.beeswarm},
    author = {Eklund, Aron and Trimble, James},
    year = {2010},
}

@article{gschwandtner_visual_2016,
    title = {Visual {Encodings} of {Temporal} {Uncertainty}: {A} {Comparative} {User} {Study}},
    volume = {22},
    doi = {10.1109/TVCG.2015.2467752},
    number = {1},
    journal = {IEEE Transactions on Visualization and Computer Graphics},
    author = {Gschwandtner, Theresia and Bögl, Markus and Federico, Paolo and Miksch, Silvia},
    year = {2016},
    pages = {539--548},
}

@article{newburger_fitting_2023,
    title = {Fitting {Bell} {Curves} to {Data} {Distributions} {Using} {Visualization}},
    volume = {29},
    doi = {10.1109/TVCG.2022.3210763},
    number = {12},
    journal = {IEEE Transactions on Visualization and Computer Graphics},
    author = {Newburger, Eric and Correll, Michael and Elmqvist, Niklas},
    year = {2023},
    pages = {5372--5383},
}

@article{kale_visual_2021,
    title = {Visual {Reasoning} {Strategies} for {Effect} {Size} {Judgments} and {Decisions}},
    volume = {27},
    doi = {10.1109/TVCG.2020.3030335},
    number = {2},
    journal = {IEEE Transactions on Visualization and Computer Graphics},
    author = {Kale, Alex and Kay, Matthew and Hullman, Jessica},
    year = {2021},
    pages = {272--282},
}

@inproceedings{fernandes_uncertainty_2018,
    address = {Montreal QC Canada},
    title = {Uncertainty {Displays} {Using} {Quantile} {Dotplots} or {CDFs} {Improve} {Transit} {Decision}-{Making}},
    doi = {10.1145/3173574.3173718},
    booktitle = {Proceedings of the 2018 {CHI} {Conference} on {Human} {Factors} in {Computing} {Systems}},
    publisher = {ACM},
    author = {Fernandes, Michael and Walls, Logan and Munson, Sean and Hullman, Jessica and Kay, Matthew},
    year = {2018},
    pages = {1--12},
}

@article{kale_hypothetical_2019,
    title = {Hypothetical {Outcome} {Plots} {Help} {Untrained} {Observers} {Judge} {Trends} in {Ambiguous} {Data}},
    volume = {25},
    doi = {10.1109/TVCG.2018.2864909},
    number = {1},
    journal = {IEEE Transactions on Visualization and Computer Graphics},
    author = {Kale, Alex and Nguyen, Francis and Kay, Matthew and Hullman, Jessica},
    year = {2019},
    pages = {892--902},
}

@article{newburger_comparing_2023,
    title = {Comparing overlapping data distributions using visualization},
    volume = {22},
    doi = {10.1177/14738716231173731},
    number = {4},
    journal = {Information Visualization},
    author = {Newburger, Eric and Elmqvist, Niklas},
    year = {2023},
    pages = {291--306},
}

@inproceedings{matejka_same_2017,
    address = {Denver Colorado USA},
    title = {Same {Stats}, {Different} {Graphs}: {Generating} {Datasets} with {Varied} {Appearance} and {Identical} {Statistics} through {Simulated} {Annealing}},
    doi = {10.1145/3025453.3025912},
    booktitle = {Proceedings of the 2017 {CHI} {Conference} on {Human} {Factors} in {Computing} {Systems}},
    author = {Matejka, Justin and Fitzmaurice, George},
    year = {2017},
    pages = {1290--1294},
}

@article{riedel_replacing_2022,
    title = {Replacing bar graphs of continuous data with more informative graphics: are we making progress?},
    volume = {136},
    doi = {10.1042/CS20220287},
    number = {15},
    journal = {Clinical Science},
    author = {Riedel, Nico and Schulz, Robert and Kazezian, Vartan and Weissgerber, Tracey},
    year = {2022},
    pages = {1139--1156},
}

@article{helske_can_2021,
    title = {Can {Visualization} {Alleviate} {Dichotomous} {Thinking}? {Effects} of {Visual} {Representations} on the {Cliff} {Effect}},
    volume = {27},
    doi = {10.1109/TVCG.2021.3073466},
    number = {8},
    journal = {IEEE Transactions on Visualization and Computer Graphics},
    author = {Helske, Jouni and Helske, Satu and Cooper, Matthew and Ynnerman, Anders and Besançon, Lonni},
    year = {2021},
    pages = {3397--3409},
}

@article{hall_professional_2022,
    title = {Professional {Differences}: {A} {Comparative} {Study} of {Visualization} {Task} {Performance} and {Spatial} {Ability} {Across} {Disciplines}},
    volume = {28},
    doi = {10.1109/TVCG.2021.3114805},
    number = {1},
    journal = {IEEE Transactions on Visualization and Computer Graphics},
    author = {Hall, Kyle Wm. and Kouroupis, Anthony and Bezerianos, Anastasia and Szafir, Danielle Albers and Collins, Christopher},
    year = {2022},
    pages = {654--664},
}

% biblatex with biber
% \printbibliography                

\end{document}